\documentclass[fleqn,usenatbib]{mnras}

\usepackage[T1]{fontenc}

\DeclareRobustCommand{\VAN}[3]{#2}
\let\VANthebibliography\thebibliography
\def\thebibliography{\DeclareRobustCommand{\VAN}[3]{##3}\VANthebibliography}

\pdfoutput=1
\usepackage{aas_macros}
\usepackage{epsfig}
\usepackage{amsmath}
\usepackage{amssymb}
\usepackage{caption}
\usepackage{leftidx}
\usepackage{natbib}
\usepackage{float}
\usepackage{enumitem}

\usepackage[T1]{fontenc}
\usepackage{lmodern}
\usepackage[rgb]{xcolor}
\usepackage{listings}
\definecolor{MyGreen}{rgb}{0.0,0.6,0.3}
\definecolor{MyPurple}{rgb}{0.6,0,0.3}
\usepackage{fancyref}
\usepackage{hyperref}
\hypersetup{colorlinks=true,citecolor=MyGreen,linkcolor=MyPurple,urlcolor=blue}

\usepackage{graphicx}	

\title[Magnetized Turbulent Mixing Layers]{Simulations of Weakly Magnetized Turbulent Mixing Layers}

\author[Zhao \& Bai]{
Xihui Zhao,$^{1}$\thanks{E-mail: zhaoxh20@mails.tsinghua.edu.cn}
Xue-Ning Bai,$^{1,2}$\thanks{E-mail: xbai@tsinghua.edu.cn}
\\
$^{1}$Institute for Advanced Study, Tsinghua University, Beijing 100084, China\\
$^{2}$Department of Astronomy, Tsinghua University, Beijing 100084, China\\
}

\date{Accepted XXX. Received YYY; in original form ZZZ}

\pubyear{2022}

\begin{document}
\label{firstpage}
\pagerange{\pageref{firstpage}--\pageref{lastpage}}
\maketitle

\begin{abstract}
Radiative turbulent mixing layers are expected to form pervasively at the phase boundaries in multiphase astrophysical systems.
This inherently small scale structure is dynamically crucial because it directly regulates the mass, momentum and energy exchanges between adjacent phases. Previous studies on hydrodynamic turbulent mixing layers have revealed the interactions between cold and hot phases in the context of the circumgalactic medium,
offering important insight into the
fate of cold clouds traveling through hot galactic winds. However, the role of magnetic field has only been sparsely investigated. 
We perform a series of 3D magnetohydrodynamics (MHD) simulations of such mixing layers in the presence of weak to modest background magnetic field.
We find that due to field amplification, even relatively weak background magnetic fields
can significantly reduce the surface brightness and inflow velocity of the hot gas in the mixing layer.
This reduction is attributed to a combination of magnetic pressure support and direct suppression of turbulent mixing, both of which alter the phase structures. Our results are largely independent of thermal conduction and converged with resolution, offering insights on the survival of cold gas in multiphase systems.
\end{abstract}




\section{Introduction}

Commonly found in astrophysical plasmas are baryons coexisting in various phases spanning a wide range of temperatures and densities. While the different phases are discrete and thermally stable on their own right, the boundaries separating them are not necessarily sharp discontinuities, but extended layers thickened by diffusive transport processes such as viscosity and thermal conduction \citep{Borkowski:90,Gnat:10}. Furthermore, turbulent motions possibly driven by Kelvin-Helmholtz instabilities (KHI) mix up different phases at the interfaces, giving rise to turbulent mixing layers (TMLs) \citep{Begelman90}. Usually, these TMLs at intermediate-temperatures radiate more efficiently thus cools more rapidly, and hence the TMLs can dominate the energetics and play an active role in shaping the phase structure. Examples include supernova remnants \citep{Kim17,Fielding18,ElBadry19}, galactic winds \citep{Gronke20a,Fielding:21,Tan23} and cosmic filaments \citep{Mandelker20}. TMLs exist on nearly all scales within and around galaxies.

In a multiphase system, understanding TMLs is crucial because it is directly through these layers that mass, momentum and energy of different phases are transported, which regulates the evolution of the system. One important venue under intense investigation is the circumgalactic medium (CGM), the gas in hot halos surrounding galaxies outside their disks or interstellar medium (ISM), but inside their viral radii on $\sim 100$ kpc scales \citep{Tumlinson:17,FGO23}. As an interface between the intergalactic medium (IGM) and galaxies, the CGM is a pivot where all components of the galactic ecosystem connect, making it a new frontier to study galaxy formation and evolution. Recent observations have revealed a variety of features in the CGM, particularly on its multiphase nature. Absorption and emission-line analysis has identified that cold dense clouds ($T\sim 10^4-10^5\rm K$) are scattered ubiquitously throughout the entire diffuse halo \citep{Hennawi15}, traveling through hot ambient gas ($T\gtrsim 10^6\rm K$) at a typical velocity of $\sim 100\rm\ km/s$. Very importantly, ultraviolet absorption lines well constrained the cool phase to have a total mass on the order of $\sim 10^9-10^{10}\ M_{\odot}$ \citep{Chen09,Chen10,Prochaska11a,Werk12,Stocke13,Stern16,Prochaska17}, indicating the CGM is a massive reservoir sustaining star formation. Therefore, a solid understanding on the origin, evolution and ultimate fate of cold clouds becomes a key element for understanding the life cycle of the galactic ecosystem. The TMLs play a crucial role here because these layers govern the energetics at cold/hot interfaces and therefore the growth or destruction of cold clouds \citep{Gronke:18,Gronke20a}. Additionally, since the TMLs reside at intermediate temperatures with higher emissivity, they provide a set of important observational diagnostics. For example, TMLs are expected to explain the high ions (such as $\rm O_{VI}$) observed in absorption spectra of high-velocity clouds around the Milky Way \citep{Savage14}.

Over the past few years, the fate of cold gas in the CGM has been extensively studied in the form of ``cloud-crushing" simulations, where the typical setup is to embed a single cold cloud ($T_{\rm cold}\sim 10^4\rm K$) in hot ambient wind ($T_{\rm hot}\sim 10^6\rm K$) with a relative speed on the order of 100 km/s. Stemmed from early hydrodynamic studies (e.g., \citealp{Klein94,Xu95}), recent cloud-crushing simulations have investigated the role of various additional physical ingredients, including thermal conduction (e.g., \cite{Bruggen16,Armillotta16}), radiative cooling \citep{Scannapieco15,Gronke:18}, magnetic fields \citep{Dursi08,McCourt15,Gronke20a,Cottle20} and cosmic rays \citep{Wiener19,Bruggen20}. However, besides the greatly expanded parameter space, these studies overall lead to diverse outcomes depending on problems setting and possibly, numerical resolution. As a result, the fate of the cold clouds, especially how various physical processes control their growth/destruction, remains elusive. We note that dynamically important TMLs are usually by necessity underresolved in cloud-scale simulations with certain outcomes dependent upon numerical schemes (e.g., \citealp{Braspenning22}), which also motivates further studies on the TMLs. As an intrinsically small-scale structure, TML can be considered as local patches at the cold-hot interfaces in the cloud-crushing problem, and the study of it helps refine cloud-scale simulations and potentially provide the necessary sub-grid physics.

The primary goal for studying the TMLs is to clarify the rate of local mass, momentum and energy exchanges between cold and hot phases. These rates are essentially encapsulated by the inflow velocity $v_{\rm in}$ of hot gas to cold gas (because on a local scale, mixing layer quickly cools down, generating more cold gas while consuming more hot gas), and is also effectively reflected in the cooling luminosity.
\cite{Begelman90} did early analytic work on TMLs. They pointed out the existence of TMLs characterized by a temperature of $T_{\rm mix}\sim \sqrt{T_{\rm cold}T_{\rm hot}}$ and width of $l_{\rm mix}\sim v_{\rm turb}t_{\rm cool}$ at the cold/hot interfaces, mediating their interactions. Here $v_{\rm turb}$ is turbulent velocity and $t_{\rm cool}$ is the cooling time scale for mixing layer to cool down to the cold phase. Their theory thus implies that $v_{\rm in}\sim l_{\rm mix}/t_{\rm cool}\sim v_{\rm turb}$. Recently, however, to investigate the abundance of high ions within TMLs, \cite{Ji:19} performed 3D plane parallel simulations of TMLs and found that $l_{\rm mix}\propto t_{\rm cool}^{1/2}$, which is inconsistent with early results. This $l_{\rm mix}\propto t_{\rm cool}^{1/2}$ scaling leads to $v_{\rm in}\propto t_{\rm cool}^{-1/2}$. Subsequently, \cite{Gronke20a} ran cloud simulations to study the growth rate of cold cloud, but derived $v_{\rm in}\propto t_{\rm cool}^{-1/4}$ from their results. This is also observed in local simulations of turbulent mixing by \citet{Mandelker20} and \citet{Fielding:20}. Later, by exploiting parallels with the turbulent combustion theory, \cite{Tan:21} reconciled the discrepancy and bridged previous results. It turns out the competition between turbulent mixing and radiative cooling is responsible for different scalings, and the dominant process largely sets the flow properties. Numerically, consistent results have also been obtained among different groups.

Thanks to the aforementioned studies of TMLs, a systematic understanding on the efficiency of turbulent mixing at small scales is being established, which offers subgrid physics for large-scale models such as galactic winds (e.g., \cite{Fielding:21,Tan23}). However, most previous works on TMLs are hydrodynamic and neglected magnetic fields. In cloud-crushing simulations, magnetic field has been shown to slow down the destruction of clouds \citep{Dursi08,Gronnow18}, but at late time the presence of magnetic fields only makes minor difference in the lifetime or mass growth of cold clouds \citep{Li20,Cottle20,Gronke20a}. Meanwhile, magnetic field geometry introduces additional complexity and
influences the morphology and acceleration of cold clouds \citep{McCourt15,BandaBarrag18,Cottle20}. On the other hand, the resolution requirement for MHD cloud-crushing simulations is also uncertain, 
due to the complex interplay between magnetic field and turbulence in the TMLs that are not necessarily properly resolved.
Therefore, we aim to quantify the role of magnetic fields by study magnetized TMLs.

In this work, we perform 3D MHD plane parallel simulations with both radiative cooling and anisotropic thermal conduction to investigate the properties of magnetized TMLs, especially in comparison with previous hydrodynamic results. We note that \cite{Ji:19} has also run a subset of 3D simulations with magnetic fields that are similar to ours, but mostly focused on the high ion abundances rather than $v_{\rm in}$. We cover a large parameter space in magnetic field and radiative cooling strength, and inspect the flow properties particularly on the inflow velocity $v_{\rm in}$ (or equivalently surface brightness $Q$, see Section \ref{sec_surfacebrightness}), the morphology and the phase distributions of magnetized TMLs. Our primary focus is on the regime with weak initial magnetization, as this branch has been sparsely investigated, and is more applicable for the plane parallel model. We find even weak initial magnetic fields (magnetic pressure $\sim 500$ times smaller than thermal pressure) can be amplified to balance thermal pressure, and substantially change the conclusions drawn from hydrodynamic simulations. Furthermore, we briefly studied the impacts of different field geometries and conductivity criteria.

This paper is structured as follows. We describe our numerical methods and implementations in Section \ref{num_method}. Then we present our results in Section \ref{3D_result}, including an overview (\ref{sub_gallery}$\sim$\ref{sub_overview}), detailed diagnostics and analysis (\ref{effects_cooling}$\sim$\ref{high_magnetization_regime}), and convergence study (\ref{sub_conv}). In Section \ref{other_parameter}, we show additional results with different magnetic field geometry and conductivity. We discuss the connections of our results to previous works and to larger scale problems, and comment on caveats in Section \ref{discussion}. We conclude in Section \ref{conclusion}.

\section{Methods}\label{num_method}

We use ATHENA++ \citep{Athena:20} to solve the following three-dimensional MHD equations on a uniform, Cartesian grid with HLLD Riemann solver:

\begin{align}
    &\frac{\partial \rho}{\partial t}+\nabla\cdot(\rho \mathbf{v})=0 \\
    &\frac{\partial }{\partial t}(\rho \mathbf{v})+\nabla\cdot(\rho \mathbf{vv}-\mathbf{BB}+\mathbf{P}^*)=0
    \\
    &\frac{\partial E}{\partial t}+\nabla\cdot[(E+P^*)\mathbf{v}-\mathbf{B}(\mathbf{B\cdot v})]+\nabla\cdot\mathbf{q}=-\varepsilon_{\rm cool}
    \\
    &\frac{\partial \mathbf{B}}{\partial t}-\nabla\times(\mathbf{v}\times\mathbf{B})=0
\end{align}
where $\rho$, $\mathbf{v}$ and $E$ are fluid density, velocity and total energy density which is defined by the sum of internal energy, kinetic energy and magnetic energy (i.e. $E\equiv P_{\rm therm}/(\gamma-1)+\rho {\rm v}^2/2+B^2/2$ where $\gamma$ is the adiabatic index). $P^*$ is total pressure including thermal $P_{\rm therm}$ and magnetic pressure $P_{\rm mag}\equiv B^2/2$, while $\mathbf{P}^*$ is the corresponding tensor. We implement thermal conduction through the term $\nabla\cdot\mathbf{q}$, which is anisotropic due to the presence of magnetic field $\mathbf{B}$. The heat flux $\mathbf{q}$ is parallel to $\mathbf{b}\equiv\mathbf{B}/\left|\mathbf{B}\right|$, and $\varepsilon_{\rm cool}$ is an external energy source representing optically thin cooling. Details about conductivity and radiative cooling are discussed below.

\subsection{Thermal Conduction}
Conductive heat flux in magnetized plasma is mostly parallel to the magnetic field since charged particles are confined around field lines. Thus we adopt an anisotropic heat flux:
\begin{align}
\mathbf{q}\equiv -\kappa_{\parallel}(\mathbf{b}\cdot\nabla T)\mathbf{b},
\end{align}
where $T$ is the temperature and $\kappa_{\parallel}$ the parallel conductivity. We implement anisotropic thermal conduction following the prescription in \cite{Sharma07}, which preserves the monotonicity in anisotropic diffusion.

The canonical Spitzer conductivity for fully ionized plasmas has been given by \cite{Spitzer:62} :
\begin{align}\label{K_Spitzer}
    \kappa_{\rm Spitzer}(T)=5.7\times 10^{-7}\ T^{5/2}\rm erg\ cm^{-1}\ s^{-1}\ K^{-1}.
\end{align}
In reality, the actual level of thermal conduction is uncertain due to small scale physics which may change the electron mean free path. Instead of using $\kappa_{\rm Spitzer}$ above, we assume a constant conductivity $\kappa_{\parallel}$ equivalent to the value of $\kappa_{\rm Spitzer}(T)$ at $T=0.8\times 10^5\ \rm K$:
\begin{align}\label{K_parallel}
    \kappa_{\parallel}=10^6\rm erg\ cm^{-1}\ s^{-1}\ K^{-1}\ ,
\end{align}
which roughly corresponds to the conductivity in the mixing layers, as also adopted in \citet{Tan:21}. We also discuss different choices of conductivity in Section \ref{field_geometry} and show that the overall results are insensitive to our choices.

For comparison purposes, we also run a set of 3D hydrodynamic simulations without magnetic fields, where conduction is isotropic with the conductivity $\kappa_{\rm iso}=\kappa_{\parallel}$.

\subsection{Radiative Cooling}
Radiative cooling is probably the most important non-ideal process in TMLs and can qualitatively alter the underlying physics \citep{Gronke:18,Fielding:21}. In our MHD equations, we add an external energy source due to radiative cooling as
\begin{align}\label{cooling}
\frac{1}{\gamma -1}\left.\frac{dP}{dt}\right|_{\rm cool}=-\varepsilon_{\rm cool}\equiv-n^2\Lambda(T)
\end{align}
where $n$ is number density and $\Lambda$ is the cooling curve as a function of temperature.

Following \cite{Fielding:20}, we specify our log-normal cooling function $\Lambda(T)$ by (i) the maximum value $\Lambda(T_{\rm mix})$, which coordinates with the cooling table calculated by \cite{Gnat:07} and (ii) the width, which is arranged so that $\Lambda(T_{\rm mix})=100\times\Lambda(T_{\rm cold/hot})$. In reality, the shape of cooling curve is more sophisticated and depends on the metallicity in the environment. Although \cite{Tan:21B} acknowledged the shape of cooling curve can change the temperature distribution, we have tested the realistic cooling table calculated by \cite{Gnat:07} and found the overall dynamics is not sensitive to the specific shape of the cooling curve, as long as it well reflects a bi-stable feature.


Cooling energy loss described by equation \ref{cooling} induces a cooling time scale:
\begin{align}\label{tcool_def}
t_{\rm cool}(T)\equiv \frac{P_{\rm therm}}{(\gamma-1)n^2\Lambda(T)}
\end{align}
Since it is the most intensive cooling within the mixing layer that concerns the system and vitalizes evolution, we refer to the cooling time of mixed $T_{\rm mix}\equiv\sqrt{T_{\rm cold}T_{\rm hot}}=1\times 10^5\rm\ K$ gas as $t_{\rm cool}$ afterwards, which is $t_{\rm cool}(T_{\rm mix})\approx 1\rm\ Myr$.


In the following numerical experiments, we adjust the strength of radiative cooling by multiplying a constant prefactor $\Lambda_0$ on the fiducial cooling function. The actual cooling function is thus given as
\begin{align}
    \Lambda(T)=\Lambda_0\Lambda_{\rm fid}(T)
\end{align}
and $\Lambda_{\rm fid}$ is the log-normal cooling function described above. By definition we have $t_{\rm cool}\propto \Lambda_0^{-1}$. Utilization of this prefactor provides the numerical convenience for studying the influence of different cooling time scales. Physically, tuning the cooling strength is equivalent to changing the ambient pressure,
where increasing $\Lambda_0$ corresponds to larger ambient pressure.

\subsection{Initialization}
In order to generate a TML, we simulate a plane-parallel shear layer, and trigger the initial KHI by imposing relative shear motion to feed turbulence that mixes up the cold and hot gas. In reality, our setup imitates a local patch at the boundary of cold clouds travelling through hot ambient gas.

We construct our 3D MHD plane parallel models closely following \cite{Ji:19,Tan:21}. The simulation domain contains $512\times128\times128$ cells, corresponding to $400\times100\times100$ pc. This means a cell length of 0.78 pc, approximately resolving the Field length $\lambda_F\equiv\sqrt{\kappa T/n^2\Lambda}$ in our simulations when thermal conduction is included \citep{Begelman&McKee90}. We arrange our coordinate system such that the $x$ axis is normal to the cold/hot interface, along which the turbulent mixing front propagates. The $y$ axis parallels shear flows, the initial relative motion feeding the KHI, and the $z$ axis is the remaining dimension. Outflow boundaries are applied in $x$ direction, and periodic boundaries to the rest. Physically, the coordinate ranges are $[-200,200]$ pc or $[-100,300]$ pc along the $x$ axis, and $[0,100]$ pc along the $y$ and $z$ axes. We use $L_{\rm box}\equiv 100\rm\ pc$ to denote the simulation box size hereafter. We adjust the bounds in $x$ direction according to cooling strength to ensure that we can capture the entire mixing layer for sufficiently long time, and in the meantime keep its evolution unaffected by the choice of boundary conditions.

We fill the negative $x$ region with cold gas $(T_{\rm cold}=10^4\rm\ K)$ and the positive $x$ region with hot gas $(T_{\rm hot}=10^6\rm\ K)$, separated by a thin $(\sim 6\ \rm cells)$ initial mixing layer $(T_{\rm mix}\equiv \sqrt{T_{\rm cold}T_{\rm hot}}=10^5\rm\ K)$ centered at $x=0$. Different phases are originally in pressure equilibrium, with number density $n_{\rm cold}=1.6\times 10^{-2}\rm cm^{-3}$ and $n_{\rm hot}=1.6\times 10^{-4}\rm cm^{-3}$. We further impose a uniform magnetic field $\mathbf{B}_0$ in the $y$ direction parallel to the initial shear flows. We choose this field orientation because in spite of the uncertainty in realistic field direction, the relative motion generally tends to rearrange the field line along the shear flows due to the frozen-in effect, thus the $\mathbf{B}_y$ component should dominate around the mixing layer. Effects of different initial field geometry will be discussed in Section \ref{field_geometry}.

The initial field strength $B_0$ is adjusted through the initial plasma beta $\beta_0\equiv P_{\rm therm,0}/P_{\rm mag,0}$, the ratio of thermal pressure to magnetic pressure. Although the value of $\beta$ is typically uncertain in real astrophysical systems, we focus on weak field regime with $50\leq\beta_0\leq 50000$ so that the background field does not prevent initial mixing, and we will show that even such weakly magnetized environments are enough to drive deviations from hydrodynamic results.

Note that by our design, $P_{\rm therm,0}$ is uniform and constant throughout all our simulations, while the initial total pressure $P_{0}^*=P_{\rm therm,0}\left(1+\frac{1}{\beta_0}\right)$ depends on $\beta_0$. But since we only choose $\beta_0\geq 50$, $P^*_{0}$ can be regarded as nearly constant. The initial sound speed in the hot gas is $c_{s,\rm hot}=\sqrt{\gamma P_{\rm therm}/\rho_{\rm hot}}=150{\rm\ km/s}$. Due to initial pressure equilibrium, we have $c_{s,\rm cold}=0.1c_{s,\rm hot}$.

We drive the initial shear motion and seed the perturbation to induce KHI using the following velocity profile:
\begin{align}
   v_y & = \frac{v_{\rm shear}}{2}{\rm tanh}\left(\frac{x}{a}\right)\\
   v_x & = \delta v\ {\rm exp}\left(-\frac{x^2}{a^2}\right){\rm cos}(k_y y){\rm cos}(k_z z)
\end{align}
where $v_{\rm shear}$ is $\rm 100\ km/s$, slightly lower than $c_{s,\rm hot}$. $a=5\ \rm pc$ is approximately the width of the initial mixing layer, and $\delta v=0.01v_{\rm shear}$. The perturbation wavelength $\lambda_{y,z}=2\pi/k_{y,z}$ equals to $L_{\rm box}$. We also apply a white noise $\sim 1\times 10^{-4}\ v_{\rm shear}$ on top of this velocity profile.

\subsection{Surface brightness}\label{sec_surfacebrightness}
In magnetized TMLs, the physical quantity of major interest is the inflow velocity $v_{\rm in}$ of hot gas flowing into the cold phase, because it directly encapsulates the the rates of mass and energy exchange between the two phases, and is a parameter that can be encoded in large-scale simulations or theories \citep{Lancaster21,Fielding:21,Tan22}. However, the inflow velocity $v_{\rm in}$, which describes the propagation of the entire turbulent mixing front, is not straightforward to measure because the simulation box can develop bulk velocities. We instead measure the surface brightness $Q$ defined by the total cooling rate:
\begin{align}\label{Q_def}
   Q\equiv \frac{1}{S}\int \varepsilon_{\rm cool}\ d {\rm v}=\frac{1}{S}\int n^2 \Lambda d {\rm v},
\end{align}
where $S=\rm 100\ pc \times 100\ pc$ is the cross-sectional area of our simulation domain.

Measuring $Q$ or $v_{\rm in}$ are roughly equivalent because in a quasi-steady state, radiative cooling energy loss should be balanced by enthalpy flux in a frame comoving with the mixing layer
\citep{Ji:19,Fielding:20,Tan:21},
\begin{align}\label{Q_vin}
   Q\approx \frac{5}{2}P v_{\rm in},
\end{align}
where $P$ is the ambient pressure surrounding the mixing layer (near the boundary at the hot end), which is approximately equal to $P_{\rm therm,0}$, as will be shown in later sections. We caution that this relationship is not rigorous, especially in high-Mach number regions \citep{Bustard22,Yang23}. In Section \ref{Q_summary} we also measure $v_{\rm in}$ and assess the relation (\ref{Q_vin}), where we found that the discrepancies are generally small.

In the following sections, we visualize the local cooling emissivity $\varepsilon_{\rm cool}=n^2\Lambda$, and normalize it by the initial values within the mixing layer:
\begin{align}
  \varepsilon_0 \equiv n_{\rm mix}^2\Lambda(T_{\rm mix}).
\end{align}
therefore $\varepsilon_0$ has linear dependence on $\Lambda_0$. To highlight the cooling radiation from mixing layers, we subtract the background cooling emission in our diagnostics below, which is defined as $n^2 \Lambda(T_{\rm cold/hot})$. We have checked that it has negligible influence to our main conclusions given our design of cooling curve.

From this point onward, we define the temperature range $2\times 10^4 {\rm K}<T<3\times 10^5{\rm K}$ to describe mixed gas in subsequent sections.

\section{3D Simulation Results}\label{3D_result}

\begin{figure*}
\begin{center}
    \includegraphics[scale=0.425]{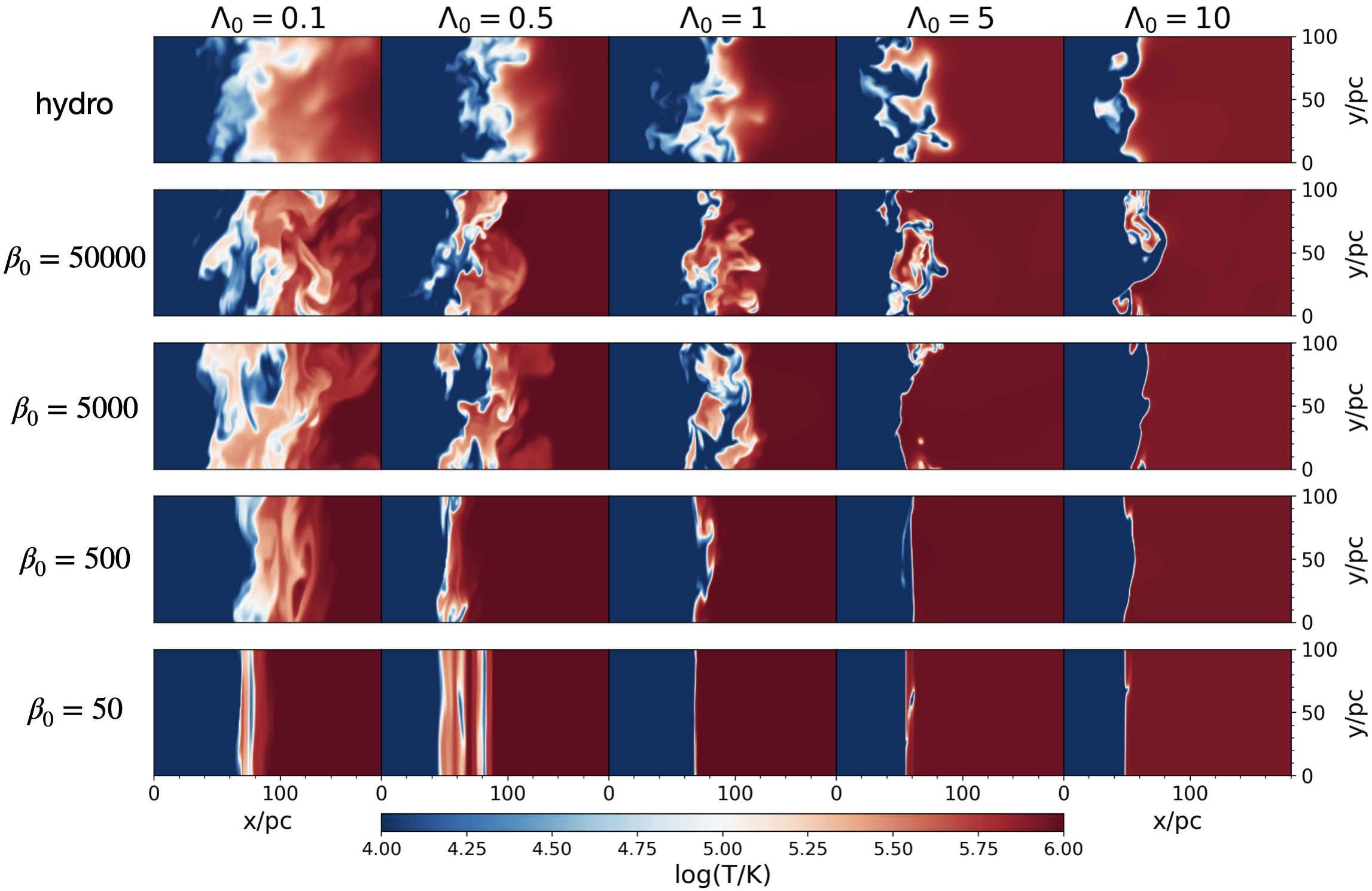}
\end{center}
    \caption{\label{gallery}Gallery of our main simulation set, which covers a parameter space with $\beta_0$ from 50000 to 50 (along with the hydrodynamic runs), and $\Lambda_0$ from 0.1 to 10. Each panel is a projected $x-y$ slice illustrating temperature fields around the mixing layers. From left to right, radiative cooling efficiency grows. From top to bottom, initial magnetic field strengthens.
    These snapshots are taken at late time during the evolution ($t>120t_{\rm shear}$ for $\beta_0\leq5000$ while $t>75t_{\rm shear}$ for $\beta_0=50000$ ), so that the systems have sufficiently developed and reached their quasi-steady states. In each panel, we have adjusted the $x-$coordinates so that the mixing layer is roughly located at the same position for all runs.
    }
\end{figure*}

\subsection{Overview of the simulation set}\label{sub_gallery}
It is useful to start by taking an overview of our simulation set on magnetized TMLs. Figure \ref{gallery} is a gallery that displays temperature slices of mixing layers seen edge on, taken from the majority of our simulations, which cover a parameter space with initial plasma $\beta_0$ from 50000 to 50 (and also hydrodynamic runs), and radiative cooling strength $\Lambda_0$ from 0.1 to 10. After the initial KHI is developed, the resulting turbulence mixes up different phases and eventually creates a quasi-steady TML. These snapshots of TMLs provide a general overview about how mixing shapes the phase distribution at the interfaces and how varying magnetic field or cooling strength affects the evolution.


We first introduce the results from hydrodynamic simulations (first row). When cooling is very inefficient $(\Lambda_0=0.1)$, we see a rather smooth temperature (and hence density) transition from the cold phase to the hot phase, showing an extended layer mostly filled with intermediate-temperature gas. This is because the long $t_{\rm cool}$ allows different phases to fully mix up well before the intermediate-temperature gas cools down, and the mixing layer is ``single-phase" \citep{Tan:21}. However, as $\Lambda_0$ increases, faster cooling shortens the lifetime of intermediate-temperature gas. Once $t_{\rm cool}$ is shorter than the mixing time scale ($t_{\rm mix}\sim L_{\rm box}/v_{\rm shear}$ which happens to occur at $\Lambda_0\approx 1$), the mixing layers lack extended intermediate-temperature regions, with abrupt jump between the hot and cold phases, giving rise to ``multi-phase" mixing layers, in which most gas is either cold or hot. The interface where a small amount of intermediate-temperature gas resides become 
corrugated and show fractal structures \citep{Fielding:20,Tan:21}.

After magnetic field is incorporated, an immediate effect is to weaken the KHI and hence turbulent mixing through magnetic tension.
We still observe an overall trend that as cooling rate increases, the phase structure in the mixing layer transitions from ``single phase" to ``multi-phase". On the other hand, the flow becomes more and more laminar as magnetic field strengthens, especially when $\beta_0\leq 500$.

In the cases with $\beta_0=50000$ (second row), the morphology of TMLs appears similar to hydrodynamic simulations (except for being more fractal presumably because anisotropic conduction is less diffusive, see also \cite{Fielding:20,Tan:21}). We checked that in such cases $Q$ (or $v_{\rm in}$) indeed only slightly deviates from hydrodynamic results (see Figure \ref{Q_scaling}). Therefore, we mainly focus on $50\leq \beta_0 \leq 5000$ in the rest of this paper.

\begin{figure*}
\begin{center}
    \includegraphics[scale=0.432]{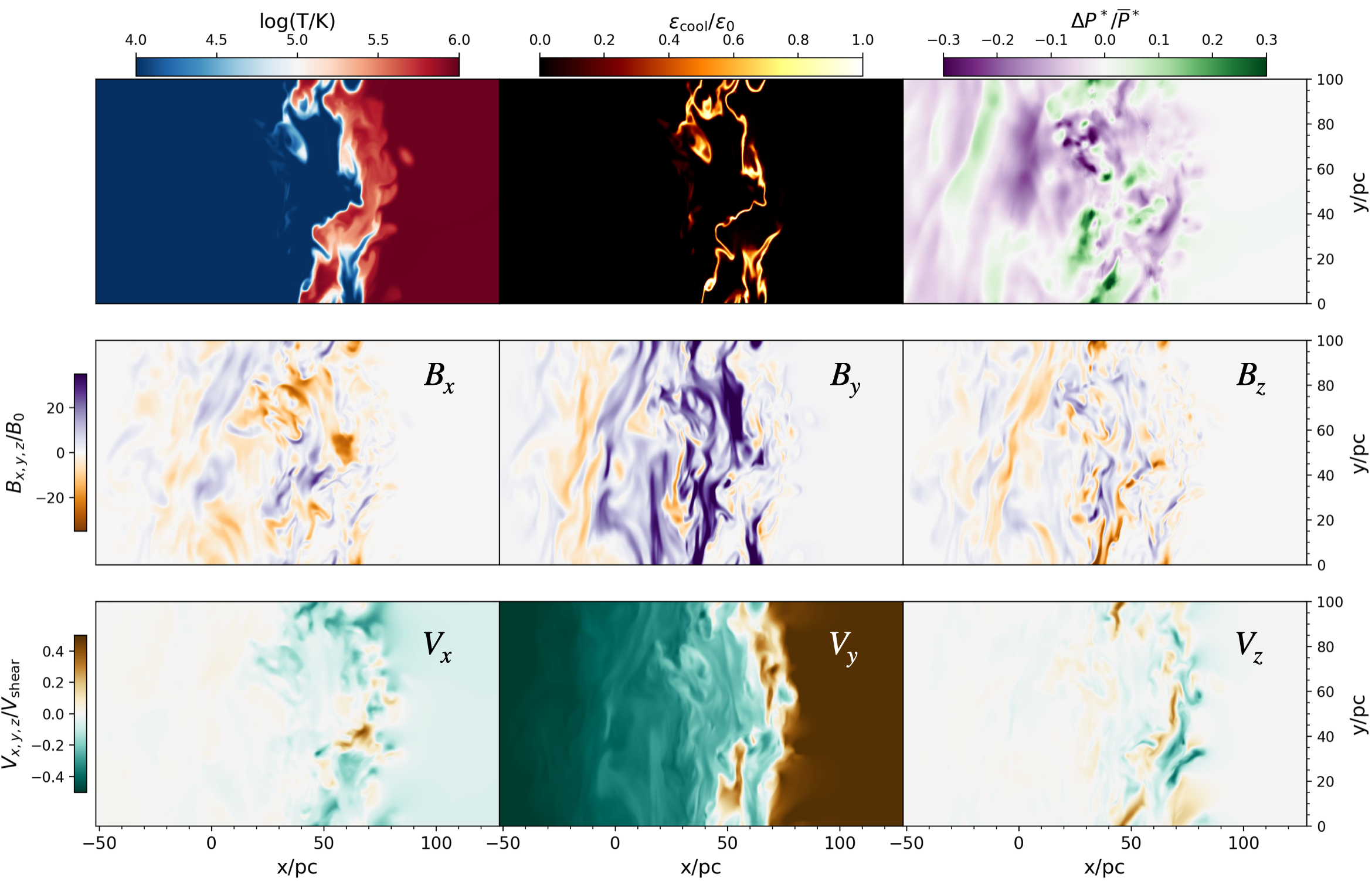}
\end{center}
    \caption{\label{overview} Properties of our fiducial magnetized TML simulation with $\beta_0=5000$, $\Lambda_0=1$ at the highest resolution $(\Delta x=L_{\rm box}/256)$, shown from
    a single slice in the $x-y$ plane, taken from the snapshot
    at $t=110t_{\rm shear}$. {\bf{Top row:}} from left to right are temperature, cooling emission and total pressure deviation. Here $P^*\equiv P_{\rm therm}+P_{\rm mag}$ and $\overline{P}^*$ is the average over the simulation box. {\bf{Middle row:}} three magnetic field components, normalized by initial strength $B_0$. {\bf{Bottom row:}} three velocity components normalized by initial shear velocity $v_{\rm shear}=100\ \rm km/s$.
    }

\end{figure*}

\subsection{General features of magnetized TMLs}\label{sub_overview}
We first take our highest resolution run and show in Figure \ref{overview} the typical features of a weakly magnetized TML. Different panels illustrate temperature, emissivity, pressure deviation, magnetic field and velocity profiles of a single slice in the $x-y$ plane. The corresponding simulation is with fiducial cooling strength $\Lambda_0=1$ and $\beta_0=5000$, a choice where magnetic fields can exert certain influence,
while also preserves the turbulence structure. This snapshot is taken at $t=110\ t_{\rm shear}$ where $t_{\rm shear}\equiv L_{\rm box}/v_{\rm shear}$ is shear timescale across the simulation box, thus turbulence has been well developed at this moment by the initial KHI.
Through Figure \ref{overview}, we introduce the main characteristics of the magnetized TML.


\begin{enumerate}[wide, labelwidth=!,itemindent=!,labelindent=0pt, leftmargin=0em, label=(\arabic*), parsep=0pt]
    \item \textit{Phase structure}

    The temperature panel shows that, instead of hosting a mixing layer that is mostly ``single-phase" as in the hydrodynamic counterpart
    (e.g., the upper-middle panel in Figure \ref{gallery}), the mixing layer in Figure \ref{overview} is more ``fractal", consisting more of discrete cold/hot phases interspersed within each other, while a small amount of mixed gas separates them. This can be understood as the suppression of mixing by magnetic field better preserves the temperature at the cold/hot phases.
    In the meantime, from the emission (middle) panel we can see that most radiation takes place within the narrow corrugated boundary where intermediate-temperature gas resides, reinforcing the more fractal nature of the mixing layer.

    \item \textit{Magnetic field amplification}
    
    During the evolution of TMLs, the initial field $\mathbf{B_0}=B_0\hat{\mathbf{y}}$ gets entangled and amplified. With weak initial field, the amplification results from the kinematic dynamo from the KHI turbulence, which is the strongest in regions around the interface between the cold and hot gas where shear is the strongest (bottom row of Figure \ref{overview}).
    In this case with $\beta_0=5000$, an amplification of $\sim 40$ times in field strength can be achieved (middle row in Figure \ref{overview}). The amplified field is highly turbulent and is dominated by the $\hat{y}-$component along the direction of shear, as expected, while the other two components also reach nearly comparable strength. This saturated field strength corresponds to a minimum $\beta \sim 3$, suggesting $P_{\rm mag}$ could be comparable to $P_{\rm therm}$ at later stage.
    In addition, note that the location of the mixing layer migrates from the initial $x=0$ toward larger $x$ as cold gas grows due to cooling. We observe that the amplified fields leave an imprint in the cold phase after the mixing layer sweeps through.
    

    
    \item \textit{Isobaric cooling?}
    
    One useful diagnostic on the dynamics of the mixing layer concerns whether this layer is isobaric (i.e., equal pressure as in the cold/hot phase).
    \cite{Ji:19} found in their hydrodynamic simulations that thermal pressure has a dip in the mixing layer that constitutes $\sim8\%$ of total pressure due to efficient cooling. This dip is compensated by turbulent pressure, reflecting that the hot gas is siphoned to the cold phase.
    \cite{Gronke20a} also identified supportive signs in their cloud-scale MHD simulations. However, \cite{Fielding:20} pointed out that pressure dips could be a signature of inadequate resolution. They suggested the system should be isobaric as long as the cooling layers are properly resolved.
    
    In our MHD simulations, where the physical resolution reaches as high as those used in \cite{Fielding:20} (although our box size is smaller), we find the answer likely mainly depends on the magnetization. In hydrodynamic cases or MHD cases with very weak $B_0$ (such as the $\Delta P^*/\bar{P}^*$ panel in Figure \ref{overview}), there are pressure deficits especially as $\Lambda_0$ increases, in agreement with \cite{Ji:19}. But when $B_0$ increases, $P_{\rm mag}$ almost fully compensates the deficits of $P_{\rm therm}$ (see Figure \ref{Pprofile_mag}),
    with negligible contribution from turbulent pressure.
    We will show when $\beta_0=500$, the situation appears isobaric quite strictly, and the results remain consistent in both our fiducial and high-resolution simulations.

\end{enumerate}

\subsection{Weakly Magnetized TMLs}\label{effects_cooling}
From here we center our discussions around the surface brightness $Q$ of magnetized TMLs, which is roughly equivalent to the hot gas inflow velocity $v_{\rm in}$. Previous hydrodynamic simulations \citep{Fielding:20,Tan:21} have shown that the value of $Q$ in the quasi-steady state is eventually determined by the balance between radiative cooling and turbulent mixing. In the following we describe how magnetic field takes part in this balance.

We start with the very weak field regime by fixing $\beta_0=5000$, where the field is sufficiently weak to respond mostly passively to gas turbulence, akin to dynamo action in the kinematic regime. We then solely adjust $\Lambda_0$ to investigate the role of cooling. In this section we also lay out the main diagnostics used throughout the rest of this paper.

\subsubsection{Time evolution of surface brightness $Q$}\label{Q_evo_cooling}

\begin{figure}
\begin{center}

\includegraphics[scale=0.415]{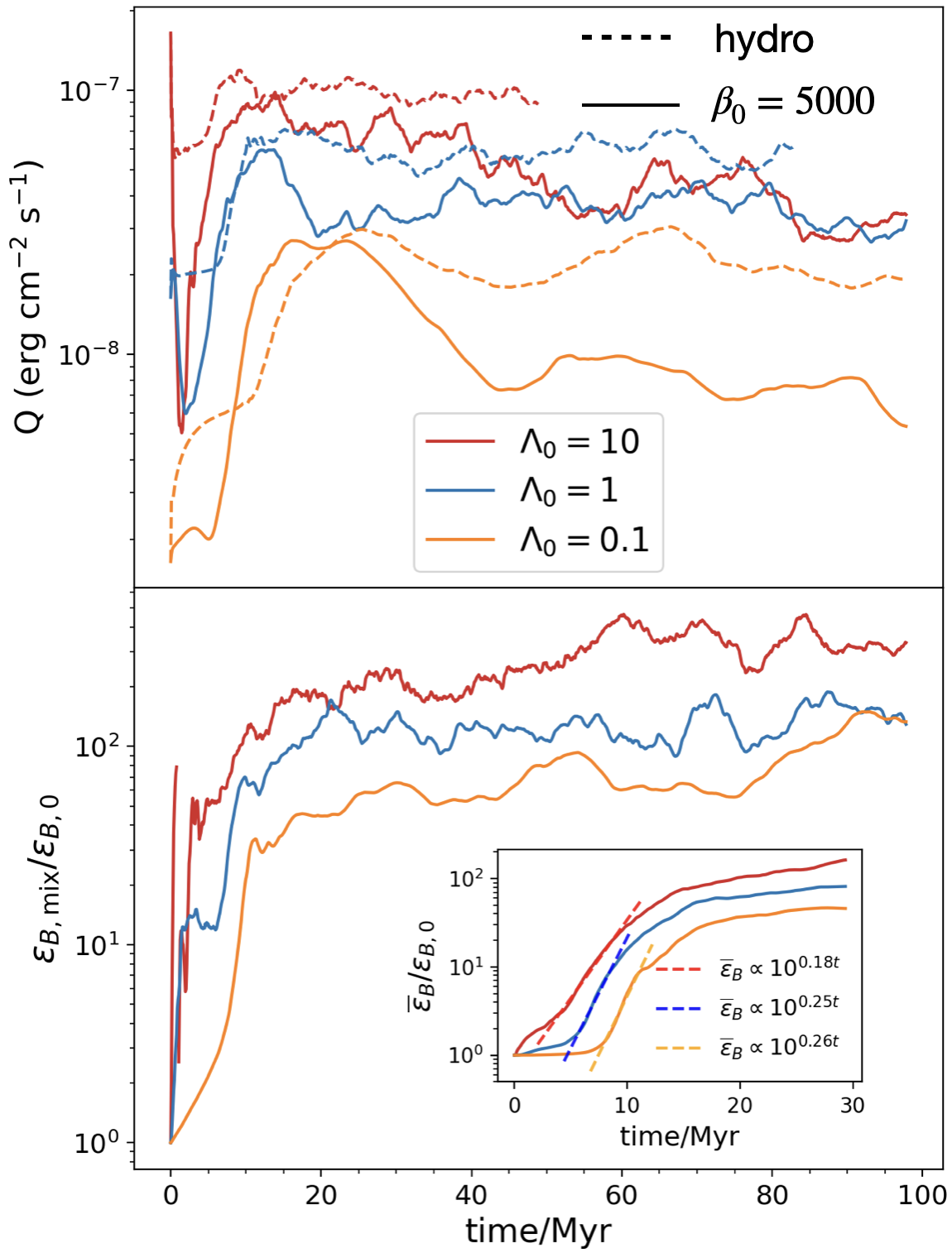}
\caption{\label{Q_cooling}Time evolution of surface brightness $Q$ (top) and average magnetic energy density $\varepsilon_{B,\rm mix}$ in the mixed gas (bottom), where $\varepsilon_{B,\rm mix}$ is normalized by initial magnetic energy density $\varepsilon_{B,0}$ (being uniform). The three colors represent weak cooling (orange), fiducial cooling (blue) and strong cooling (red) regimes, respectively. Solid lines are from MHD simulations with $\beta_0=5000$, in comparison with dashed lines from hydrodynamic simulations. Red and blue dashed lines are cut off at $t\approx 50 \rm Myr$ and $85\rm Myr$ as the hot gas gets fully exhausted. The inset in the bottom panel shows the corresponding evolution of magnetic energy density averaged in a region that contains TMLs during the first $30\rm Myr$. At early stage $\overline{\varepsilon}_B$ undergoes exponential growth fitted by dashed lines.
}

\end{center}
\end{figure}

 The top panel in Figure \ref{Q_cooling} shows the time evolution of $Q$ in different cooling regimes with $\beta_0=5000$ (solid lines), in comparison with hydrodynamic results (dashed lines). Note that we cut off red and blue dashed lines at $t\approx 50\rm Myr$ and $t\approx 85\rm Myr$, because afterwards hot gas is exhausted, leaving a box full of cold gas.
 
 At early stage ($t\lesssim 10\rm Myr$), $Q$ grows rapidly in all cases since turbulence fed by the initial KHI mixes up cold and hot gas, generating the intermediate phase with strong emission. However, after reaching the peak values, the two sets of simulations start to differ: instead of maintaining a steady level as in hydrodynamic cases, $Q$ in the MHD cases drops by a factor $\gtrsim 2$.
 
 This distinguished behavior can be intuitively understood through the bottom panel, where we show the evolution of average magnetic densities $\varepsilon_{B,\rm mix}$ in the mixed regions. To calculate $\varepsilon_{B,\rm mix}$, we average $B^2/B_0^2$ within the mixed gas defined by the temperature range $2\times 10^4{\rm K}<T<3\times 10^5{\rm K}$ (gas out of this temperature range has negligible contribution to $Q$). In the beginning, magnetic fields get quickly intensified by compression (i.e., as hot gas cools into cold gas) and/or turbulent mixing, initiating the rapid growth of $\varepsilon_{B, \rm mix}$, which closely correlates with the initial increasing stage of $Q$. After $\varepsilon_{B, \rm mix}$ ends its rapid growth, it slowly increases/becomes steady as $Q$ gradually decreases/maintains a steady value, which suggests amplified magnetic fields in turn suppress further mixing and hence cooling emission, and this is accompanied by reduced $Q$ values. We also observed the exponential growth, followed by flattening (saturation) in the evolution of $\overline{\varepsilon}_B$ (magnetic energy density averaged in the region that contains TMLs for the first $30\rm Myr$), as indicated in the inset of the bottom panel. This is indicative of a turbulent dynamo (e.g., \cite{Brandenburg05,Federrath16}) from the development of the KHI turbulence, though further study is needed to better characterize its properties. From the figure it is clear that more efficient cooling eventually leads to stronger magnetic field amplification.

Altogether, we think the key process in magnetized TMLs is to establish a balance between magnetic field amplification induced by cooling and/or turbulence, followed by magnetic suppression of mixing and hence cooling emission. Therefore, even weakly magnetized environment can possibly make a major difference on the properties of the mixing layers when cooling is very strong.

\begin{figure}
    \includegraphics[scale=0.245]{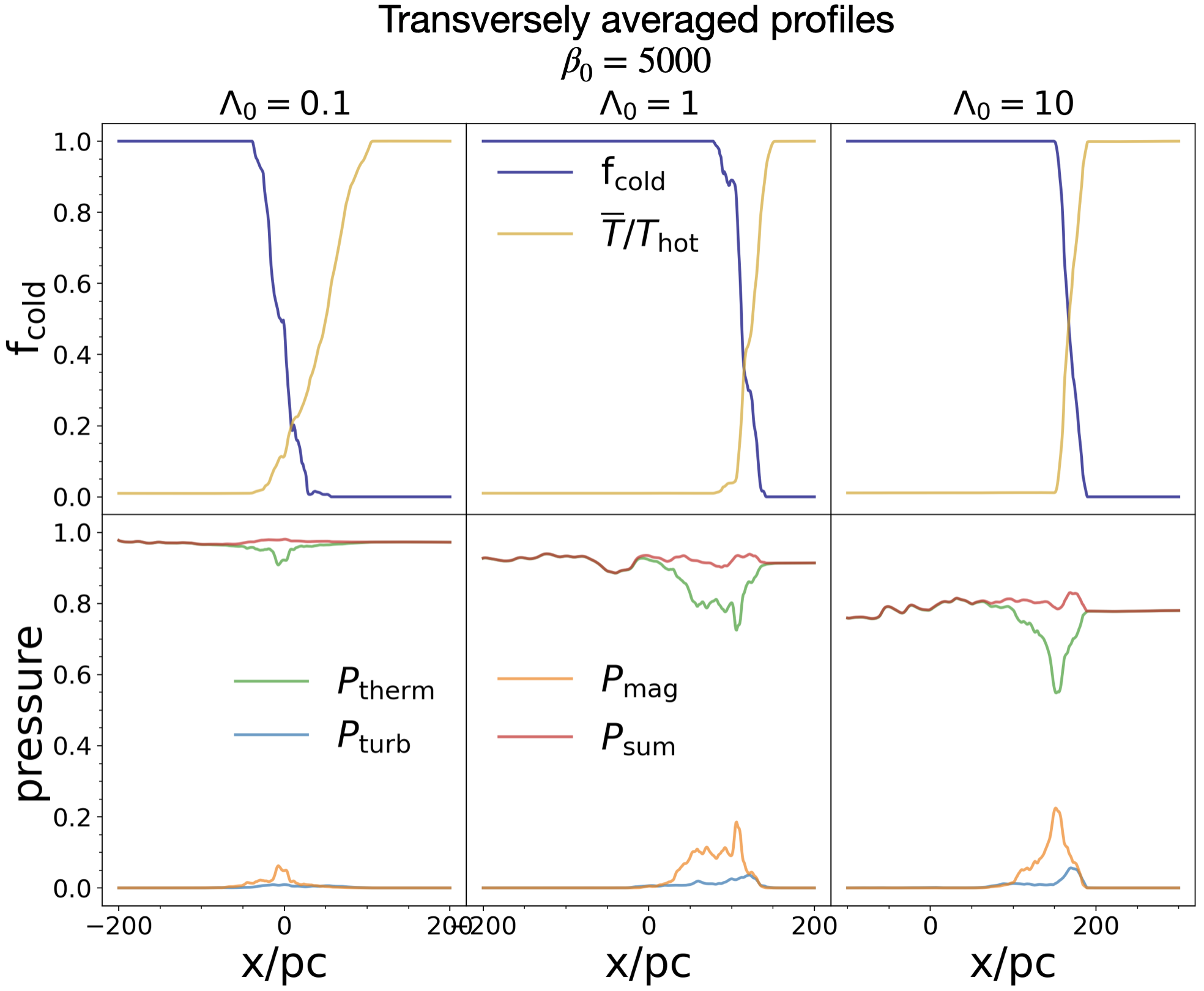}
    \caption{\label{Pprofile_cooling}Profiles of transversely averaged physical quantities along the $x$ direction. All mixing layers are initially located at $x=0$. {\bf{Top}}: Temperature and volumetric fraction of cold gas (defined as $T<5\times 10^4\ \rm K$). {\bf{Bottom}}: Thermal, magnetic, turbulent pressure and their sum, normalized by the initial total pressure $P^*_0$ in each case.}
\end{figure}

\subsubsection{Profiles and morphology of mixing layers}

Next we analyze the differences between magnetized and hydrodynamic TMLs in detail. In Figure \ref{Pprofile_cooling} we draw the profiles of transversely averaged temperatures, volumetric fractions of cold gas and pressure at the late stages in each simulation. From the top row, we clearly see that the width of the mixing layers narrows as $\Lambda_0$ increases. This is likely because faster cooling naturally inhibits the existence of intermedaite-temperature gas. Besides the layer width, \cite{Tan:21} has pointed out that the average temperature of a transverse slice should follow the simple estimate $\overline{T}\approx {\rm f}_{\rm cold}T_{\rm cold}+(1-{\rm f}_{\rm cold})T_{\rm hot}$ when mixing layers are
fractal
with little intermediate-temperature gas. This further suggests $\overline{T}/T_{\rm hot}+{\rm f_{cold}}\approx 1$ since $T_{\rm hot}\gg T_{\rm cold}$. Indeed, we see when $\Lambda_0=10$ the profile of $\overline{T}/T_{\rm hot}$ well tracks $\rm f_{\rm cold}$, while they deviate in the $\Lambda_0=0.1$ case because there is abundant volume-filling intermediate-temperature gas.

In bottom panels, we see that the final level of $P_{\rm mag}$ amplification increases with increasing cooling rate, in accordance with the findings in Figure \ref{Q_cooling}. In spite of the same value of $\beta_0$, $P_{\rm mag}$ remains negligible during the entire evolution of weakly cooling TMLs, while fast cooling can eventually lead to a $P_{\rm mag}\sim 0.5 P_{\rm therm}$. We also point out the peaks of $P_{\rm mag}$ are tracked by cold gas adjacent to mixing layers. This suggests the strongest magnetization is usually achieved in the cold phases which underwent turbulent mixing, rather than in the mixed region where field amplification is currently taking place (also see central panels in Figure \ref{overview}). We additionally calculate the turbulent pressure $P_{\rm turb}\equiv \left<\rho \delta v_x^2\right>$, and find $P_{\rm turb}$ is almost negligible even
for such large $\beta_0$ (compare with hydro case in Figure \ref{Pprofile_mag}). Here, the total pressure $P_{\rm sum}$ is largely flat (isobaric) while still fluctuates. On the other hand, we will see when $\beta_0 \lesssim 500$, the mixing layer is almost strictly isobaric.

\begin{figure}
\begin{center}

\includegraphics[scale=0.391]{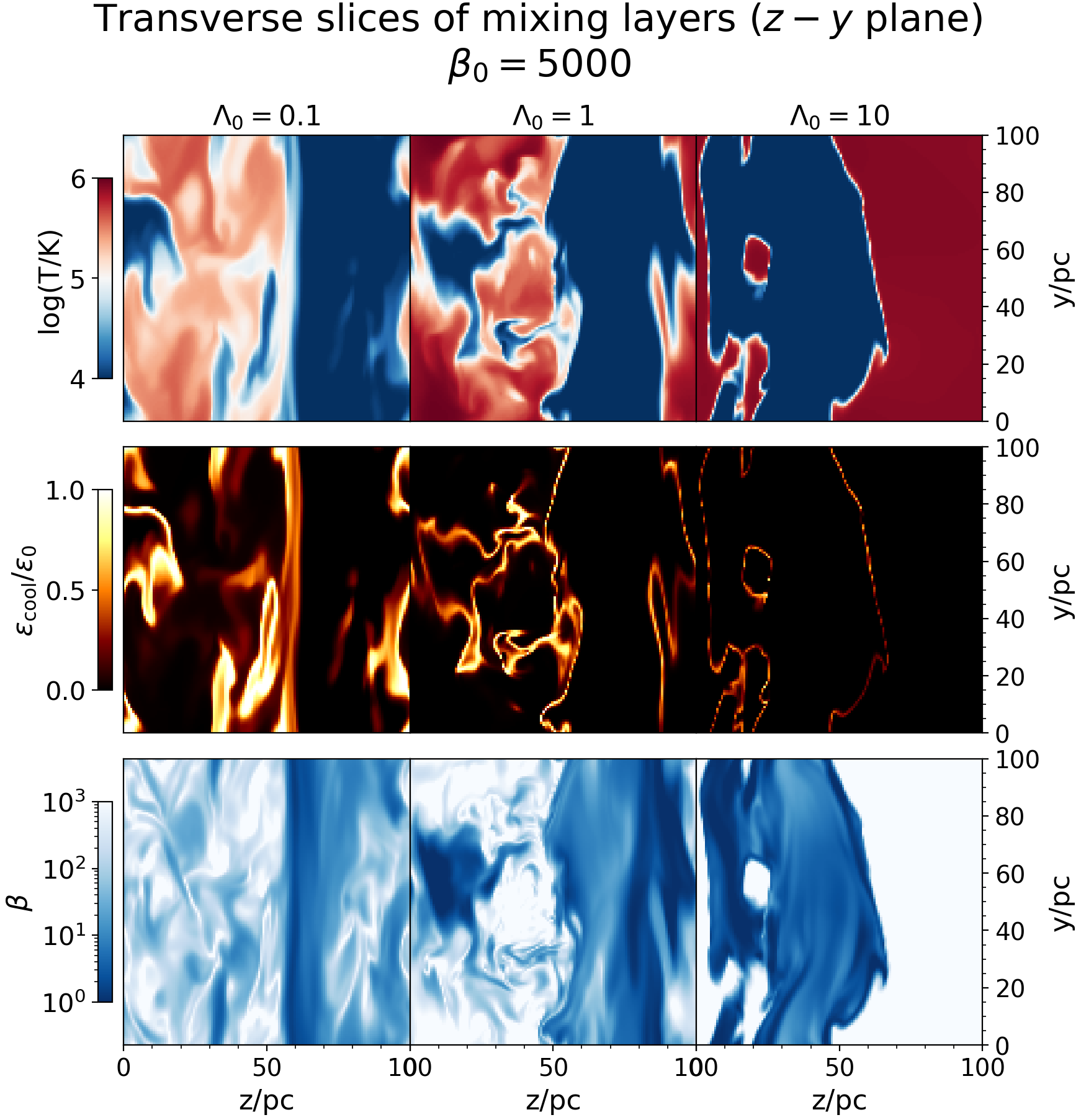}
\caption{\label{section_cooling}Face-on views of mixing layers, taken at the position where volumetric fraction of cold gas $\rm f_{cold}=0.5$ in Figure \ref{Pprofile_cooling}. These $z-y$ slices are from simulations with $\beta_0=5000$, and $\Lambda_0$ increases from left to right. {\bf Top row:} temperature. {\bf Middle row:} cooling emission normalized by initial cooling energy loss rates $\varepsilon_0$ within mixing layers (which has linear dependence on $\Lambda_0$). {\bf Bottom row:} plasma $\beta$, which illustrates strong $P_{\rm mag}$ mostly emerges in cold phase.}

\end{center}
\end{figure}

In Figure \ref{section_cooling} we visualize the transverse intersections of the TML at the positions where $\rm f_{cold}=0.5$. At first glance, there is obviously less intermediate-temperature gas as $\Lambda_0$ increases, which is also evinced from the cooling emission panels (in the middle row). As cooling strengthens from left to right, besides the area of bright regions shrinks, we see when $\Lambda_0=10$ there is hardly any bright emission even in gas with $T=10^5\ \rm K$ where the cooling rate $\Lambda(T)$ peaks. Since our normalization has already taken into account the prefactor $\Lambda_0$, such a dark pattern implies reduced contribution solely from the $n^2$ term in $\varepsilon_{\rm cool}$, which is consistent with the deficit in thermal pressure sustained by the amplified $P_{\rm mag}$. The bottom row in Figure \ref{section_cooling} indeed reveals a rising level of magnetization as cooling becomes more efficient, and stronger cooling makes the boundary separating weakly/strongly magnetized regions more distinct.

\subsubsection{Density and temperature distributions within the mixing layers}\label{subsub_PDF}

According to the definition of surface brightness $Q$ (equation \ref{Q_def}), its value mathematically only depends on two aspects:\begin{enumerate}[wide, labelwidth=!,itemindent=!,labelindent=0pt, leftmargin=0em, label=(\arabic*), parsep=0pt]
    \item  $n_T$, average density of gas with temperature around $T$
    \item $V_T$, volume occupied by gas with temperature around $T$, or equivalently the probability density function (PDF) of temperature in our simulations.
\end{enumerate}
With the above information at hand, we can then directly estimate 
\begin{equation}\label{Q_est}
Q\approx \sum\limits_T n^2_T\Lambda(T) V_T
\end{equation}

Physically, (1) and (2) correspond to magnetic pressure support and magnetic suppression of turbulent mixing, respectively. Note that in hydrodynamic simulations, (1) is a marginal factor because the deficits of $P_{\rm therm}$, even if they exist, are very minor. Therefore, the isobaric relation $n_T\propto T^{-1}$ roughly holds. In MHD, on the other hand, both (1) and (2) can significantly deviate from hydrodynamic results.

\begin{figure}
\begin{center}

\includegraphics[scale=0.33]{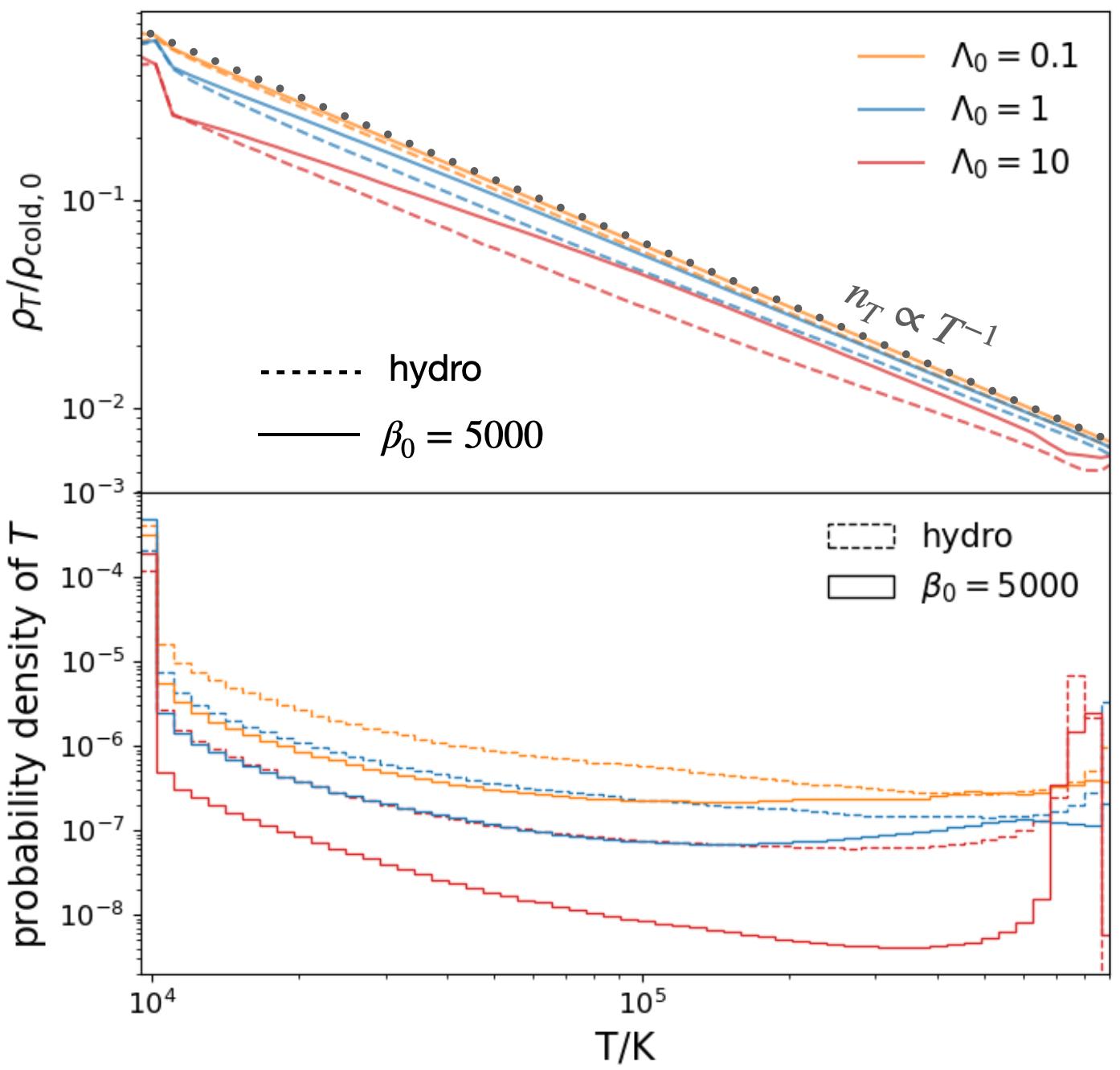}
\caption{\label{hist_cooling}{\bf{Top:}} Mean gas densities averaged over 60 temperature bins measured from both hydrodynamic (dashed line) and MHD simulations with $\beta_0=5000$ (solid line) in different cooling regimes. Grey dotted line indicates the relation $n_T\propto T^{-1}$, which should approximately hold if $P_{\rm therm}$ is in equilibrium. {\bf{Bottom:}} PDFs of temperature measured from the same set of simulations above. Two panels share the same set of temperature bins. In making each line, we average over snapshots from $t=70\rm Myr$ to the end of our simulation, or the moment right before hot gas is completely consumed.}

\end{center}
\end{figure}



To isolate magnetic influences on each aspect, we draw the density distributions as a function of temperature $\rho_T$ and the temperature PDF in Figure \ref{hist_cooling}. The top panel shows averaged densities in individual temperature bins together with the temperature PDF in the bottom panel. Note that for all cases shown, an $n_T\propto T^{-1}$ relation approximately hold, indicating $P_{\rm therm}$ is largely constant (in equilibrium).
For the three different cooling strengths, the MHD cases only show slight deviations from their hydrodynamic counterparts by a factor of $<2$.
In fact, $\rho_T$ (hence $P_{\rm therm}$) is even higher in MHD simulations when $\beta_0=5000$, likely because magnetic fields here are too weak to create substantial $P_{\rm therm}$ deficit in the mixing layer, but can
suppress the overall cooling rate (also compare the middle column in Figure \ref{Pprofile_cooling} with the left column in Figure \ref{Pprofile_mag}).

In the bottom panel, the temperature PDFs peak at $10^4$K and $10^6$K by design, while their distributions at intermediate temperatures reflect the mixing efficiency.
We see that for weak cooling ($\Lambda_0=0.1$),
the temperature PDF in the 
MHD case is only modestly suppressed compared with the hydrodynamic counterpart, consistent with our earlier discussion.
In the $\Lambda_0=10$ case, however, it can be reduced by an order of magnitude (i.e., $V_T$ $\sim 0.1$ of that in hydrodynamic case when $\Lambda_0=10$), suggesting that $Q$ can be reduced by the significant suppression on turbulent mixing when cooling is strong, which is also in line with stronger magnetic field amplification (Figure \ref{Q_evo_cooling}).

To summarize the results from
weakly magnetized TMLs $(\beta_0\sim 5000)$,
their properties are largely
degenerate to hydrodynamic TMLs when cooling is inefficient ($\Lambda_0=0.1$). As cooling intensifies, there is progressively stronger magnetic field amplification. Although $P_{\rm mag}$ is yet to reach equipartition to offset $P_{\rm therm}$, it effectively leads to suppression of mixing and reduction in $Q$.


\subsection{Modestly magnetized TMLs}\label{high_magnetization_regime}
We now turn to cases where initial magnetic fields are mildly stronger ($\beta_0=500\ \rm and\ 50$). We demonstrate that in such situations, magnetized and hydrodynamic TMLs differ in most aspects, including surface brightness $Q$, mixing layer morphology, density distribution and temperature PDF.

\subsubsection{Reduced Surface brightness $Q$}\label{Q_summary}
\begin{figure*}
\begin{center}

\includegraphics[scale=0.408]{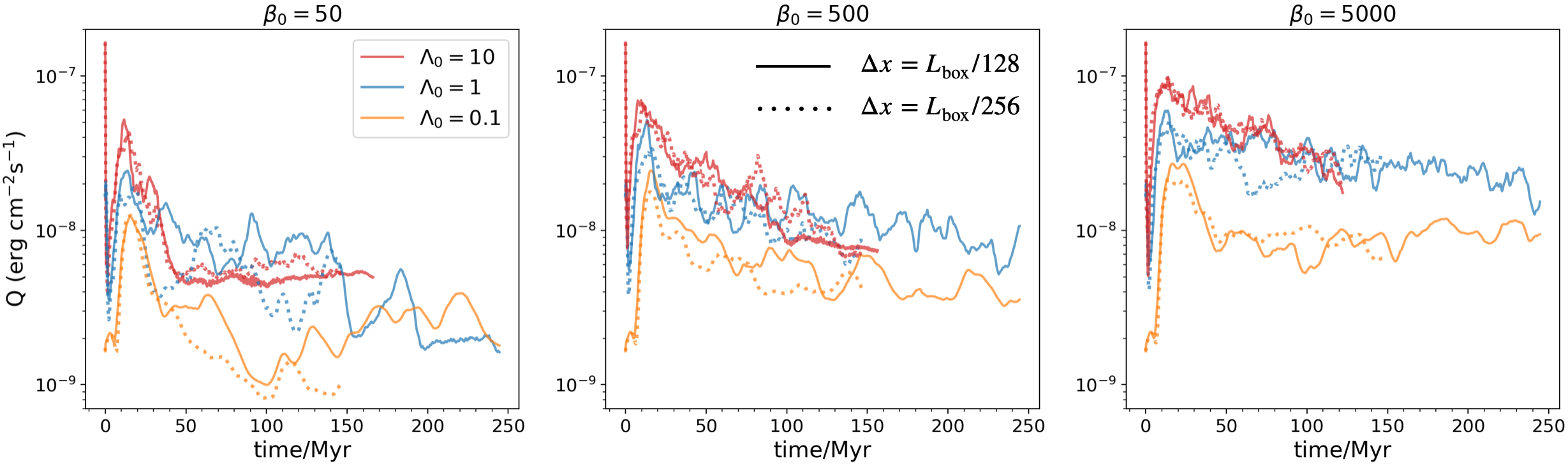}
\caption{\label{Q_conv}Time evolution of surface brightness $Q$ similar to Fig. \ref{Q_cooling}, but with $\beta_0$ ranging from $50$ to $5000$, and also includes higher resolution results. Titles explain the initial plasma beta. Solid lines are from fiducial runs $(\Delta x=L_{\rm box}/128)$ while dotted lines represent doubled resolution $(\Delta x=L_{\rm box}/256)$. Some lines end halfway as the hot gas get completely exhausted in such cases, and we only run $150{\rm Myr}$ for doubled resolution runs.}

\end{center}
\end{figure*}

We again start from the time evolution of $Q$ in magnetized TMLs. From Figure \ref{Q_conv}, we see oscillatory but persistent decline of $Q$ after peaking at $t\sim 10$ Myr in most MHD cases, similar to the discussion on Figure \ref{Q_cooling}, and there are generally larger fluctuations when stronger magnetic fields are involved. With stronger magnetization, $Q$ is consistently reduced in all cooling regimes,
and at late periods, the $Q$ values become similar regardless of cooling strength.



\begin{figure*}
\begin{center}

\includegraphics[scale=0.425]{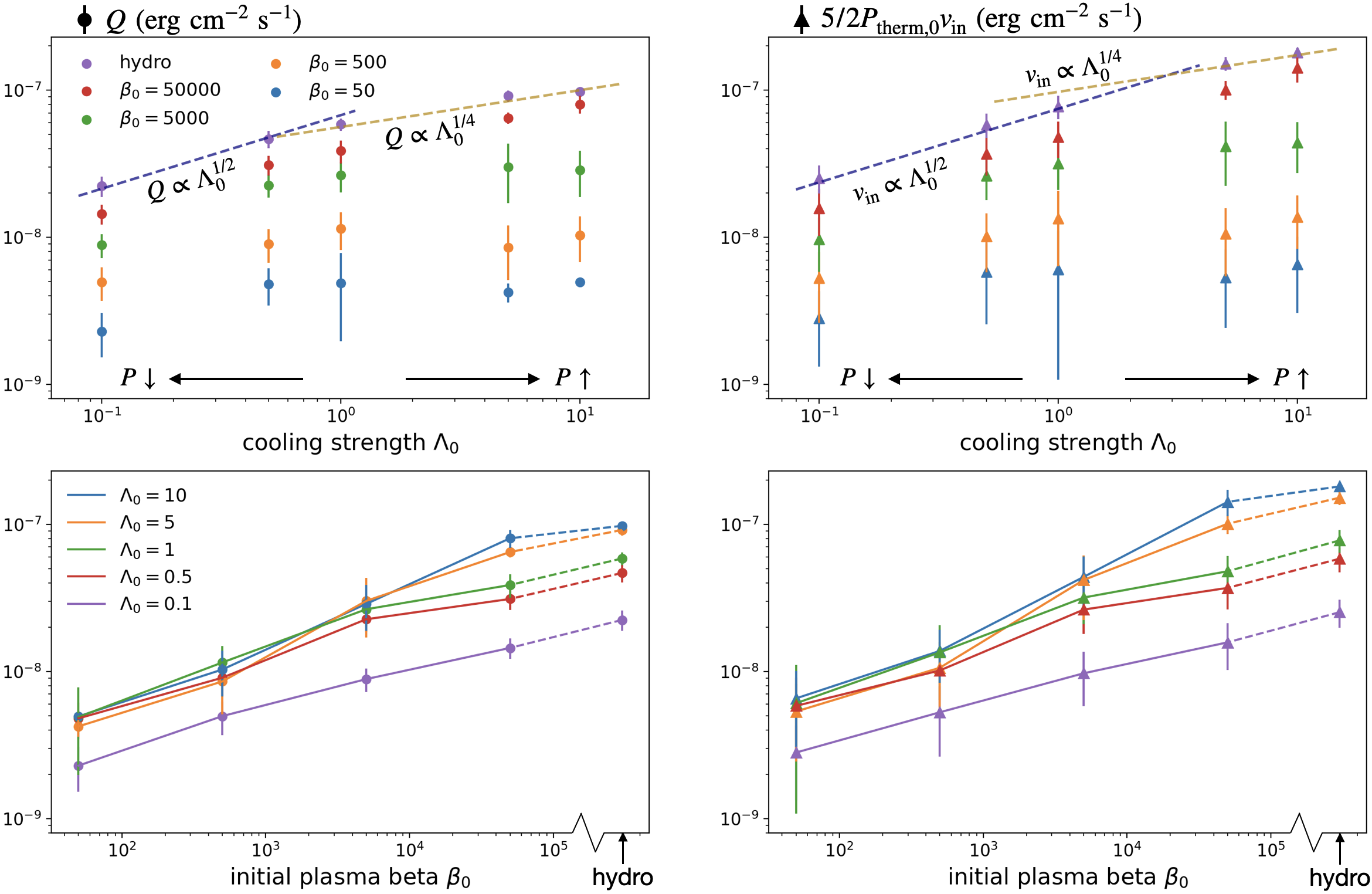}
\caption{\label{Q_scaling}{\bf Top row:} Average surface brightness (left column) and inflow velocity of hot gas (right column) as functions of cooling strength $\Lambda_0$. Numerically, tuning the cooling strength is equivalent to changing the pressure. {\bf Bottom row:} Same data points, but are drawn as functions of $\beta_0$. For each data point, the time average starts at $70\ \rm Myr$, and ends when the simulation terminates $(\geq 200\ \rm Myr)$ or the hot gas is going to be consumed. We also quote $1\sigma$ error bars reflecting fluctuation levels. The two dashed lines in the top row indicate the two scalings $\propto \Lambda_0^{1/2}$ and $\propto \Lambda_0^{1/4}$, respectively.}

\end{center}
\end{figure*}

To illustrate how $Q$ is affected by cooling strength and magnetization, we calculate averaged $Q$ in each of our simulation runs and draw the dependence on $\Lambda_0$ (top left panel) and $\beta_0$ (bottom left panel) in Figure \ref{Q_scaling}. Since evolution of $Q$ in MHD is much more fluctuating than the hydrodynamic case and sometimes shows long-term trends instead of being steady, here we simply assert that the time average always starts at $t_{\rm start}=50\rm Myr$, and ends when our simulations terminate $(t_{\rm stop}\geq 200\rm Myr)$, or hot gas is completely exhausted. This can be justified given the typical timescale at global scale (e.g., cloud-crushing timescale in the CGM) is on the order of a few 10Myrs \citep{Scannapieco15,Li20}. We quote the 1$\sigma$ error bars reflecting the uncertainties from such fluctuations.

Parallel to $Q$, the growth of the cold phase is directly related to $v_{\rm in}$, which is expected to be proportional to $Q$ through Equation (\ref{Q_vin}), although corrections may apply due to magnetic fields and turbulence.
We measure $v_{\rm in}$ by firstly recording the positions of TMLs as x-coordinate average of mixed gas $(2\times 10^4 {\rm K}<T<3\times 10^5 {\rm K})$, then calculate the propagation speed of TMLs and subtract it with the average $v_x$ of the hot boundary.
In the right column of Figure \ref{Q_scaling}, we show the same scaling relation but for $v_{\rm in}$.
To facilitate comparison,
we
multiply $v_{\rm in}$ by a constant $5/2P_{\rm therm,0}$ (background thermal pressure, which is largely unchanged over time) so that the equivalence between $Q$ and $v_{\rm in}$ is directly assessed.
We do observe there is some difference between $Q$ and $5/2P_{\rm therm,0}v_{\rm in}$, where the latter tends to be systematically smaller. This is mainly due to additional contributions from turbulence ad magnetic fields. On the other hand, the difference is within a factor of $2$, and
generally speaking,
the scaling relation measured by $v_{\rm in}$ remains identical to that measured in $Q$.


The purple dots in the top panels of Figure \ref{Q_scaling} represent our hydrodynamic reproductions of \cite{Tan:21}. The data well match the two-piece scaling $Q \propto t_{\rm cool}^{-1/2}$ and $Q \propto t_{\rm cool}^{-1/4}$ for weak and strong cooling, respectively. \cite{Tan:21} successfully explained these relations by exploiting the parallels between the turbulent mixing layers and the combustion theory, where the hot gas is the ``fuel" that ``burns" (i.e. cools radiatively). They borrowed the dimensionless Damkhöler number \citep{Damk:40}:
\begin{equation}\label{Da_eq}
    {\rm Da}\equiv \frac{L_{\rm box}}{u' t_{\rm cool}}
\end{equation}
which is the ratio of turbulence eddy turnover time scale to the cooling time scale. Here $u'$ is the turbulent velocity. If we estimate $u'\sim v_{\rm shear}=100{\rm km/s}$, and plug in $L_{\rm box}=100{\rm pc},\ t_{\rm cool}\approx 1{\rm Myr}\ (\Lambda_0=1)$, then ${\rm Da}\approx 0.98$ for our fiducial cooling case.
According to their analogy, the mixing fronts are ``single-phase" or ``multi-phase" in the ${\rm Da}<1$ and ${\rm Da}>1$ regimes (called ``laminar flame" and ``turbulent flame'' in the combustion literature), respectively, and are thus subject to different scaling relations.
Note that \cite{Fielding:20} also explained $Q\propto t_{\rm cool}^{-1/4}$ from the perspective of fractal dimensions, which we will discuss in Section \ref{frac_theory}.

However, while the properties of hydrodynamic TMLs can be well described by theory, the presence of even weak magnetic fields immediately complicates the situation. We see relatively large fluctuations in $Q$ in most cases with $\beta_0\leq 500$, with no clear sign of any specific functional relation between $Q$ and $\Lambda_0$.
The trend of an increasing $Q$ with $\Lambda_0$ is significantly attenuated under higher levels of magnetization, as also discussed earlier, and the specific role of ${\rm Da}=1$ setting the regime transition is no longer applicable. In fact, reading from the bottom panels of Figure \ref{Q_scaling}, it appears that the value of $Q$ (and $v_{\rm in}$) converges towards stronger magnetization within the uncertainties as long as $\Lambda_0\geq 0.5$. The weak cooling branch $(\Lambda_0=0.1)$, however, seems somewhat isolated from other cases. This is another fact that suggests some coupling between cooling and magnetic fields could be at play in magnetized TMLs. From the figure we can see $Q$ can be suppressed by an order of magnitude when $\beta_0\lesssim 500$. Therefore, the rate of local energy exchange between cold and hot phases is significantly restricted even in relatively weakly magnetized environments.

\subsubsection{Isobaric profile and elongated morphology of mixing layers}

\begin{figure}
    \includegraphics[scale=0.283]{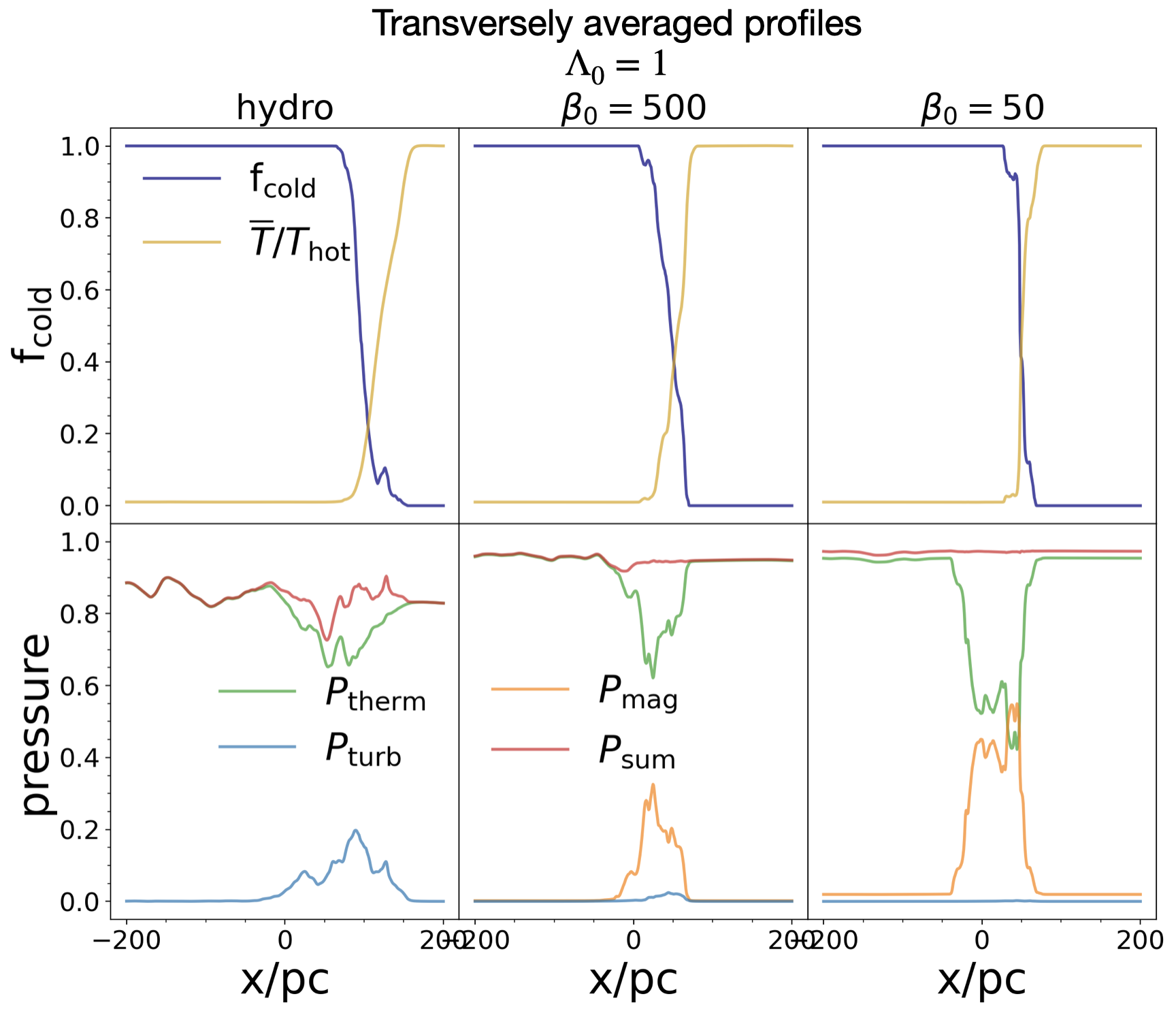}
    \caption{\label{Pprofile_mag}Profiles of transversely averaged physical quantities along $x$ direction, similar to Fig. \ref{Pprofile_cooling} but here we fix $\Lambda_0$ while adjust magnetization. All mixing layers are initially located at $x=0$. $\bf{Top}$: Temperature and volumetric fraction of cold gas (defined as $T<5\times 10^4\ \rm K$). $\bf{Bottom}$: Thermal, magnetic, turbulent pressure and their sum, separately normalized by initial total pressure $P^*_0$. When $\beta_0\lesssim 500$, the total pressure is in well equilibrium while the turbulent pressure is negligible.
    }
\end{figure}

To examine how initial field strength affects the states of magnetized TMLs, we first show in Figure \ref{Pprofile_mag} the transversely averaged profiles of temperature and pressure. We see that when $\beta_0\leq 500$, $P_{\rm mag}$ is more easily amplified to a value close to or exceeding $P_{\rm therm}$, and magnetic tension then stabilizes the initial KHI and inhibits gas mixing. Therefore, the temperature gradients become steeper as $\beta_0$ decreases. Probably due to the suppressed turbulence, profiles of total pressure are flat in a very strict manner, compared with hydrodynamic results (left column) or $\beta_0=5000$ cases in Figure \ref{Pprofile_cooling}. While it is debatable whether cooling is isobaric in the pure hydrodynamic case, we see that when $\beta_0\lesssim 500$, turbulent pressure should be negligible and hot gas is unlikely to be siphoned by total pressure cavities. In addition, the strong $P_{\rm mag}$ observed here could potentially explain the $P_{\rm therm}$ imbalance identified in the CGM observed by the Hubble Space Telescope \citep{Werk14}.

\begin{figure}
\begin{center}

\includegraphics[scale=0.385]{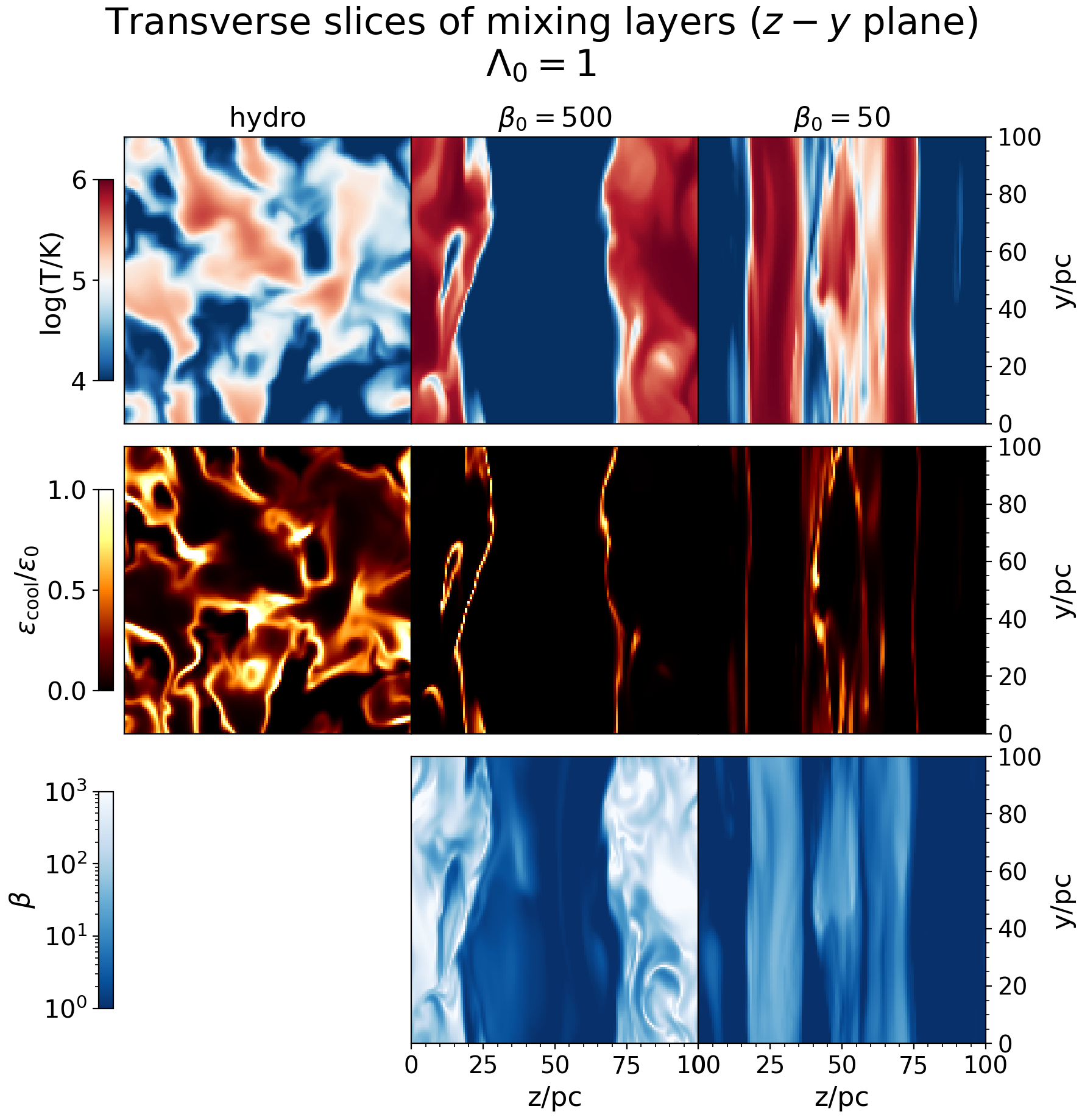}
\caption{\label{section_mag}Transverse slices of mixing layers, taken at the position where volumetric fraction of cold gas $\rm f_{cold}=0.5$ in Figure \ref{Pprofile_mag}. Slices are taken from simulations with $\Lambda_0=1$, but from left to right are simulations with $\beta_0=\infty$ (hydro), 500 and 50, respectively. {\bf Top row:} temperature. {\bf Middle row:} cooling emission normalized by initial cooling energy loss rates $\varepsilon_0$ within the mixing layers (which has linear dependence on $\Lambda_0$). {\bf Bottom row:} plasma $\beta$, which illustrates that magnetization is the strongest in the cold phases.}

\end{center}
\end{figure}

We also notice that a different morphology of mixing layers gradually emerge when $\beta_0\lesssim 500$. In fact, from Figure \ref{gallery}, we have already seen the flows become quite laminar in the bottom two rows, presumably because of the constraints by stronger magnetic tension. Besides the laminar flows seen edge on, we report that the transverse slices of mixing layers in Figure \ref{section_mag} also demonstrate a much less turbulent pattern. From the top panels in Figure \ref{section_mag}, we observe the formation of elongated stripes along the $y$ direction (i.e., background magnetic field) in the MHD runs, where the hot/cold gas are interspersed with each other. From the bottom panels, we see that strong magnetizations $\beta \lesssim 1$ are ubiquitously realized in cold phases. Partly due to enhanced magnetic pressure, the emissivity in these strips becomes weaker as magnetization increases.

We note that \cite{Ji:19} have run MHD plane parallel simulations similar to ours, and they also found shear amplified magnetic fields stabilize the turbulence at the interfaces through magnetic tension, generating almost laminar flows (see their Figure 11).

\subsubsection{Density distributions and temperature PDFs}\label{subsub_PDF_mag}

\begin{figure*}
\begin{center}
\includegraphics[scale=0.47]{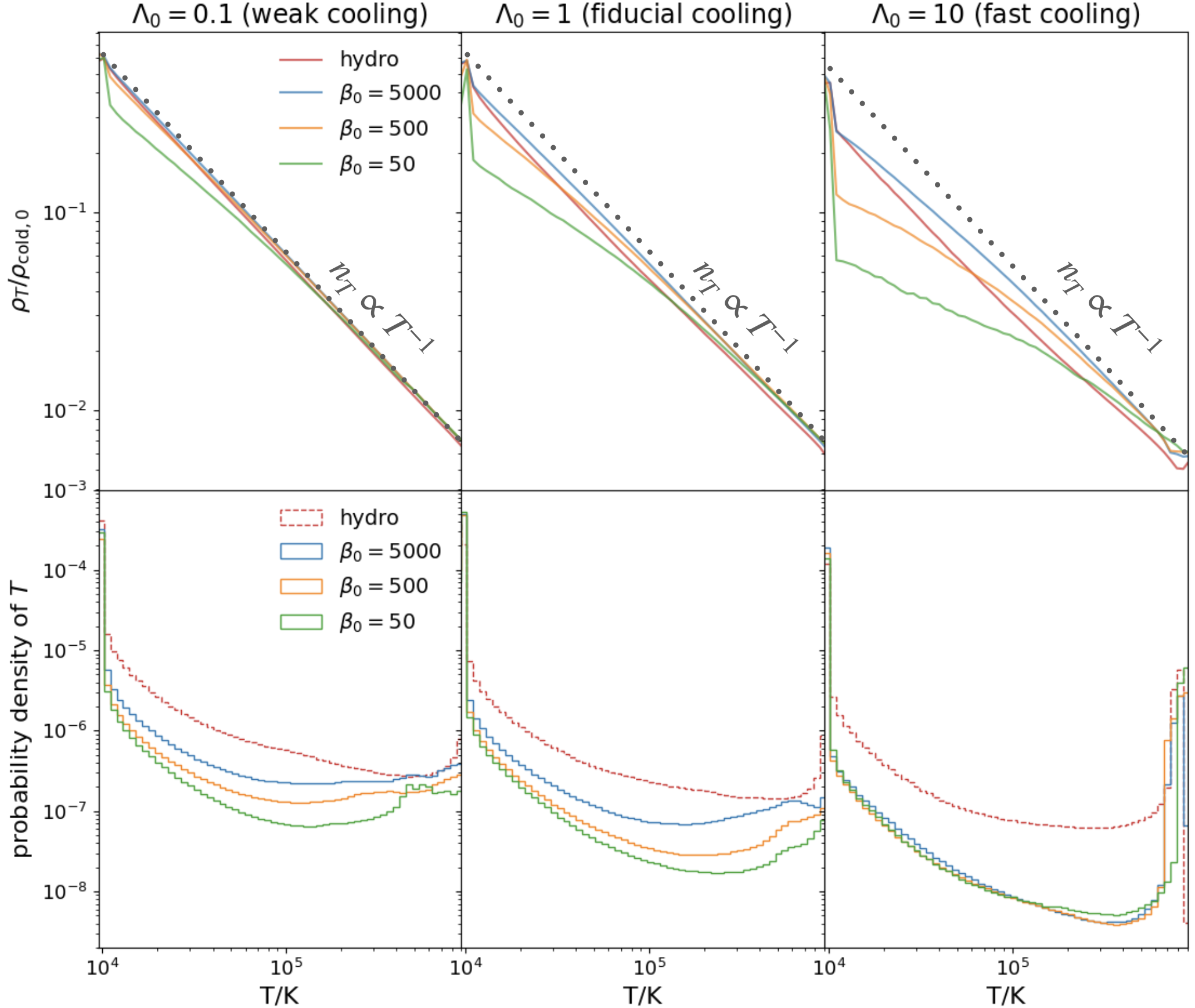}
\caption{\label{hist_mag}{\bf{Top:}} Mean density distributions as a function of temperature. {\bf{Bottom:}} Temperature PDFs. In making each line, we average over all snapshots from $t\geq 70$ Myr to the end of the simulation (same as in Figure \ref{Q_scaling}). From the left to right columns, $\Lambda_0$ increases from weak cooling to strong cooling. Different colors represent different initial magnetizations. Grey dotted lines show the $n_T\propto T^{-1}$ relation which should hold when $P_{\rm therm}$ is in equilibrium.
}

\end{center}
\end{figure*}

Following the discussion about equation \ref{Q_est} and the spirit in Section \ref{subsub_PDF}, we then isolate what exactly caused the decrease of surface brightness in each case, density deficit sustained by $P_{\rm mag}$ or suppression of mixing which prevents the generation of intermediate temperature gas? We again draw density distributions on temperature and temperature PDFs in Figure \ref{hist_mag} to isolate the two parts.

Recall that in the weak field limit (see Figure \ref{hist_cooling} where $\beta_0=5000$), $Q$ is mostly depressed by the reduction of intermediate-temperature gas, instead of creating density cavities sustained by $P_{\rm mag}$. But as we investigate systems with larger initial $B_0$, we see a different pattern in Figure \ref{hist_mag}: In the weak cooling regime ($\Lambda_0=0.1$), the situation is quite similar to Figure \ref{hist_cooling}, where density distribution barely deviates from the $n_T\propto T^{-1}$ relation. As $B_0$ increases, there is a higher level of suppression on temperature PDF and hence the depression of $Q$. However, in the strong cooling regime ($\Lambda_0=10$), although MHD temperature PDFs are also substantially suppressed compared to the hydrodynamic result, they remarkably overlap with each other regardless of $\beta_0$. On the other hand, the density contrasts among cases with different $\beta_0$ are magnified. As $B_0$ increases, regions with strong magnetization not just mainly reside in the cold phase, but also permeate towards the hot phase, sustaining significant density deficits in a broader temperature range. Therefore, when cooling is fast, the major effect of stronger magnetic fields is to increase $P_{\rm mag}$ within the intermediate phase, instead of further suppression of turbulent mixing.
These results suggest that there is a limit at which turbulent mixing can be suppressed (achieved with fastest cooling), beyond which further reduction in $Q$ is through magnetic field amplification.

\subsection{Convergence}\label{sub_conv}

To assess the robustness of our simulation results, we further conduct convergence tests for our 3D MHD simulations doubling the resolution ($\Delta x=L_{\rm box}/256$). We show the results on the evolution of $Q$ also in Figure \ref{Q_conv} by dotted lines. Generally speaking, in terms of surface brightness $Q$, our simulations reaches convergence, where the evolution of $Q$ values between the fiducial and high-resolution runs typically closely follow each other.
We note that similar convergence behavior was found in earlier hydrodynamic and MHD simulations of \cite{Ji:19} and hydrodynamic simulations of \cite{Tan:21}.

Besides the overall dynamics, we find density distributions and temperature PDFs are also well converged, and we show relevant diagnostics in Appendix \ref{Appendix_conv} to avoid distraction.
Additionally, it is important to assess whether large scale results can faithfully reflect the local energy exchange between cold and hot gas, by running simulations with worse resolutions. We again leave the results to Appendix \ref{Appendix_conv}, but report here that at later stage, $Q$ appear largely converged when $\Delta x=L_{\rm box}/32$ (a factor of 4 coarser than our fiducial resolution). Overall, this supports the reliability of simulation results at more global (e.g., cloud) scale. We also caution that magnetization is usually stronger in cloud simulations (usually $\beta \lesssim 1$), while our plane parallel setup may suffer from unrealistic boundary conditions when adopting such strong magnetic fields.

\section{Other freedom in parameter space}\label{other_parameter}

As an initial study, our parameter space simply covers two dimensions, initial magnetization $\beta_0$ and cooling strength $\Lambda_0$. However, many other physical properties may influence the state of radiative magnetized TMLs, such as magnetic field geometry and conductivity. In this section we show results from our tests adopting different initial field orientations and conductivity prescriptions. We leave detailed investigation and exploration on other effects to future works.

\subsection{Magnetic field geometry}\label{field_geometry}

\begin{figure}
\begin{center}

\includegraphics[scale=0.24]{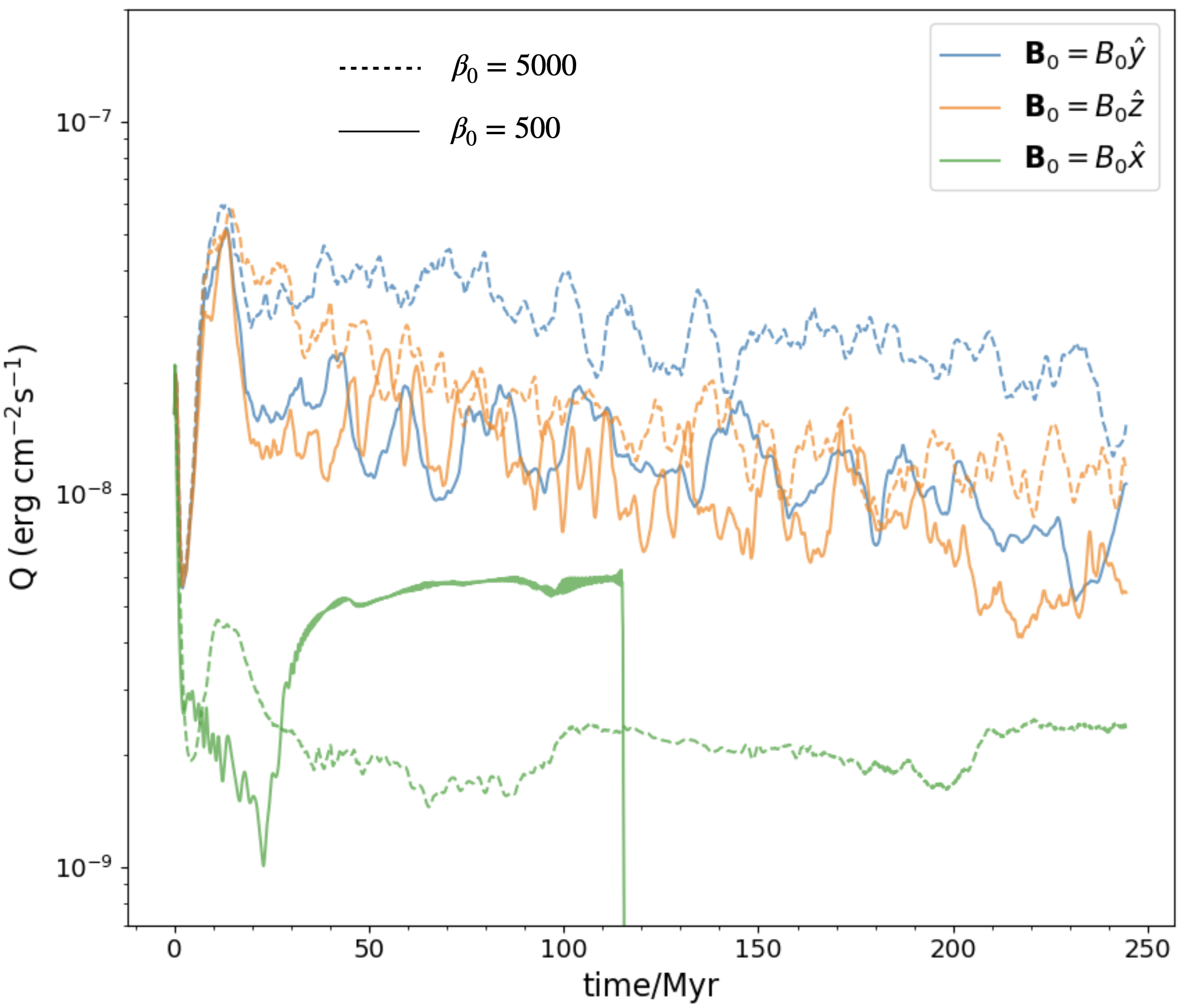}
\caption{\label{Q_orient}Time evolution of surface brightness $Q$ in simulations with different initial field orientations. Dashed lines are in the weak field limit $(\beta_0=5000)$, while solid lines indicate modestly weak fields $(\beta_0=500)$. We choose $\Lambda_0=1$ for all the cases shown here. Results are similar when $\mathbf{B}_0$ is parallel to the initial interface (blue and orange lines). However, once $\mathbf{B}_0$ is normal to the mixing front (along $\hat{x}$), it strongly suppresses $Q$ (green dashed line), or quickly shear amplifies the field to push cold gas out of the domain (green solid line).}

\end{center}
\end{figure}

Previous studies on adiabatic MHD KHI have suggested the importance of geometry and amplitude of the initial magnetic field $\mathbf{B_0}$, both theoretically \citep{Chandra:61,Miura:82} and numerically \citep{Jones:97,Ryu:20,Ji:19}. We thus investigate how different $\mathbf{B_0}$ orientations affect $Q$.
Figure \ref{Q_orient} displays time evolution of $Q$ in simulations with $\mathbf{B_0}$ along three separate axes. In these runs, we fix $\Lambda_0=1$ and examine two different initial field strengths, $\beta_0=5000$ and $500$. Our original results discussed in previous sections are shown in blue lines for benchmark, and below we discuss the results from the two remaining field orientations.

\begin{enumerate}[wide, labelwidth=!,itemindent=!,labelindent=0pt, leftmargin=0em, parsep=0pt]

    \item $\mathbf{B_0}=B_0 \hat{\mathbf{z}}$
    
    The orange lines in Figure \ref{Q_orient} stand for results from the simulations with $\mathbf{B_0}$ along the $\hat{z}$ direction, i.e., perpendicular to both shear flow direction and the interface normal. The comparable values of orange lines and benchmark blue lines indicate similar influences on stabilizing mixing layers, except $Q$ appears to have a weaker dependence on $\beta_0$ in the $\mathbf{B}_0$ along $\hat{z}$ cases.
    On the other hand, this initial field direction is not expected to interfere with the development of the adiabatic KHI,
    and without cooling, the flow from the simulations is indeed essentially hydrodynamic \citep{Ji:19}.
    Once cooling is on, \cite{Ji:19} found that mixing is suppressed to the same level as cases with $\mathbf{B_0}$ parallel to shear flows, in agreement with our results. This fact again stresses the importance of radiative cooling in the problem.

    We also checked the density distributions and temperature PDFs in this sets of simulations, and found no substantial difference from Figure \ref{hist_mag}, further reinforcing the similarities between the $\mathbf{B_0}=B_0 \hat{\mathbf{y}}$ and $\mathbf{B_0}=B_0\hat{\mathbf{z}}$ cases.

    \item $\mathbf{B_0}=B_0 \hat{\mathbf{x}}$
    
   The green lines in Figure \ref{Q_orient} denote results from the simulations with $\mathbf{B_0}$ along $\hat{x}$, i.e., normal to the cold/hot interface. Compared with other two field orientations, an obvious feature is the much stronger suppression of $Q$. Since the initial field lines are both perpendicular to shear flows and normal to the interface, they get continuously twisted by the shear flows into the $\hat{y}$ direction (along the shear) since the very beginning, giving rise to quick field amplification resulting in the suppression of the initial KHI. Consequently, there is no background equilibrium state, and there is only very limited mixing \citep{Ji:19}.

   The behavior is even more peculiar when $\beta_0=500$, in which case we see a steep decrease of $Q$ around $t\sim 115\rm Myr$. We found the reason is that, magnetic fields get shear amplified so fast that a large $P_{\rm mag}$ quickly builds up in the hot phase. Then the large pressure in hot phase soon pushes the cold gas away, leaving a box full of hot gas with small cooling radiation. This scenario probably implies a possibility that at global scale, the cold gas can be squeezed by the rapid shear amplification of the field in the hot phase. However, it is perhaps more likely that this result is an artifact of our local simulation setup, and this scenario should be better studied in a global simulation setting.

\end{enumerate}

Our choices for field geometry are far from exhausting all possibilities.
In particular, tangled initial magnetic configuration has been considered in studying the global cloud-crushing problem (e.g. \cite{McCourt15,BB17}), which lead to different consequences on the cloud clumping factor, filament morphology and other related observable quantities. Given the localized nature of the TMLs, our study may still be considered as a part of the global problem, and offers useful benchmark on the role of magnetic fields at local scales. In the meantime, we acknowledge that the role of more complex global field geometry on turbulent mixing deserves further study.

\subsection{Conductivity}

Thermal conduction can hinder the onset of hydrodynamic instabilities in the cloud-crushing problem, and has been known to change cloud morphology \citep{Bruggen16}, accelerate the evaporation of small clouds \citep{Cowie77} but substantially prolong the lifetime of large clouds \citep{Armillotta16,Li20}. However, the
effect of thermal conduction
around the local mixing layers is yet uncertain, especially with the presence of magnetic fields.

Previous hydrodynamic TML works with (isotropic) constant conductivity (e.g. \cite{Tan:21}) have shown that $Q$ is insensitive to conduction, as long as turbulent diffusion dominates heat transport. On the other hand \cite{Tan:21B} found that a temperature-dependent conductivity such as Spitzer conductivity \citep{Spitzer:62} can cause substantial difference in the temperature distribution and elemental column density within the mixing layers. Here, we briefly assess consequences of different conductivity prescriptions, which would help better constrain its consequences in global-scale models (e.g., cloud growth criterion \citep{Gronke20a} and galactic winds \citep{Fielding:21,Tan23}).

All simulations presented above adopt a constant anisotropic conductivity (equation \ref{K_parallel}). In the following, we compare them with MHD simulations that employ $\kappa_{\rm Spitzer}$ (equation \ref{K_Spitzer}), which is better physically motivated. Considering that electron transport may be hindered by micro-scale instabilities that are not well understood, the canonical Spitzer value is likely suppressed by certain factor, often taken to be an order of magnitude (e.g., \citealp{RC16,Komarov18,Drake21,Meinecke22}). Therefore, we also test the cases with conductivity being $0.1\kappa_{\rm Spitzer}$.

\begin{figure}
\begin{center}

\includegraphics[scale=0.25]{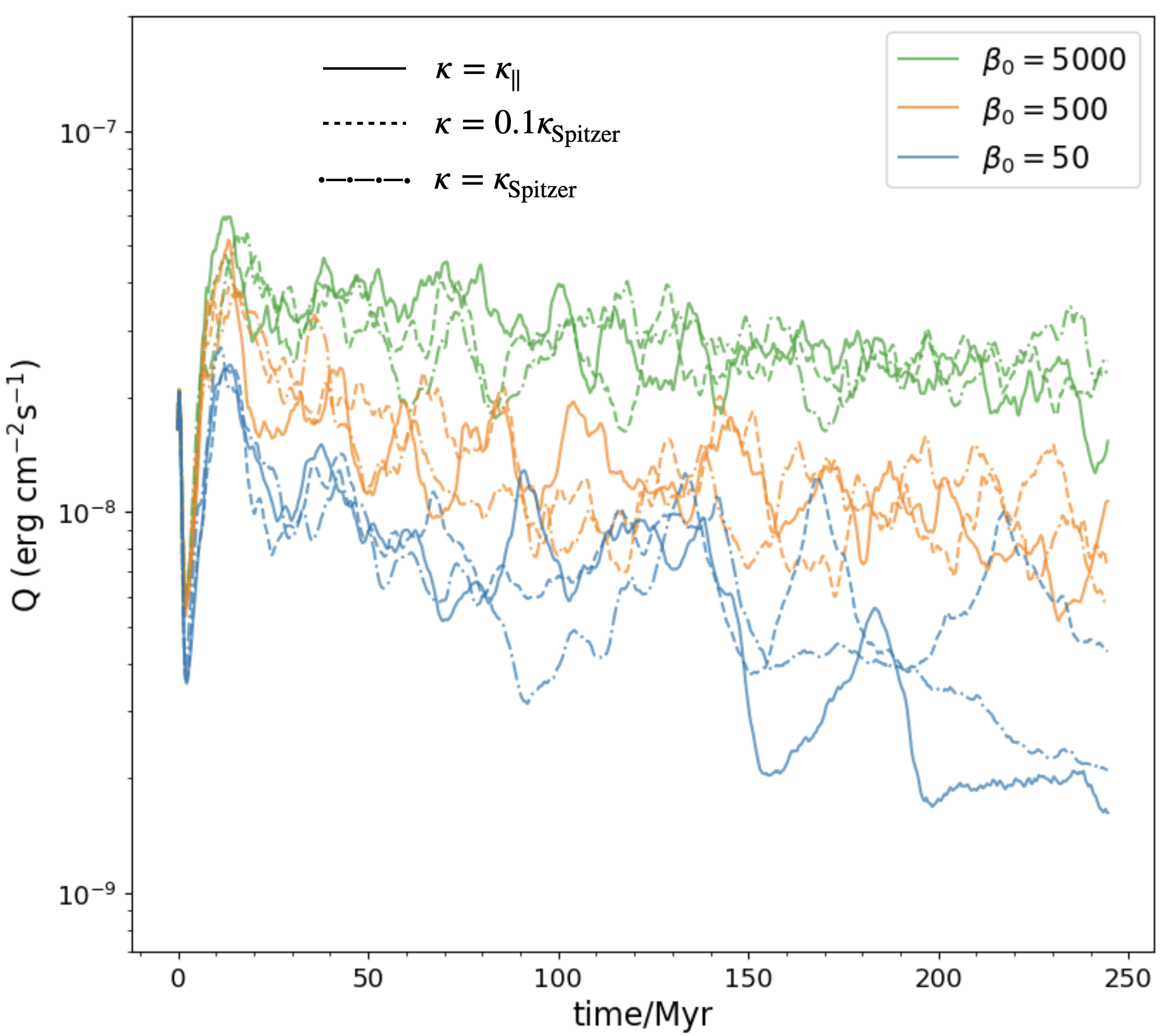}
\caption{\label{Q_Ksp}Time evolution of surface brightness $Q$, in simulations with different conductivity prescriptions, with $\Lambda_0=1$ for all the cases shown here. Three colors represent different initial level of magnetization. Different types of conductivities are indicated by line style. Generally speaking, evolution of $Q$ is not sensitive to conductivity.}

\end{center}
\end{figure}

We show in Figure \ref{Q_Ksp} the evolution of $Q$ in cases with different initial magnetization and conductivity prescriptions, while we fix $\Lambda_0=1$.
We see that the curves for different conductivity prescriptions closely follow each other, regardless of initial field strength. This result suggests that different conductivity prescriptions
do not strongly affect the energy exchange rate between cold and hot gas. The reason is likely due to the dominance of the $y$ component magnetic fields (see Figure \ref{overview}), which is perpendicular to the overall temperature gradient. In fact, we indeed observe (though not shown) that, using constant conductivity, the mean heat flux across the TML in MHD simulations is more than an order of magnitude smaller than that in the hydrodynamic simulations. Even using the Spitzer conductivity (which enhances heat transport), the resulting mean heat flux in the MHD simulations is no more than that in hydrodynamic simulations adopting constant conductivity. Recently \cite{Bruggen23} also observed magnetic fields strongly limit anisotropic thermal conduction in their cloud simulation, which is consistent with our picture here.

\begin{figure*}
\begin{center}
\includegraphics[scale=0.38]{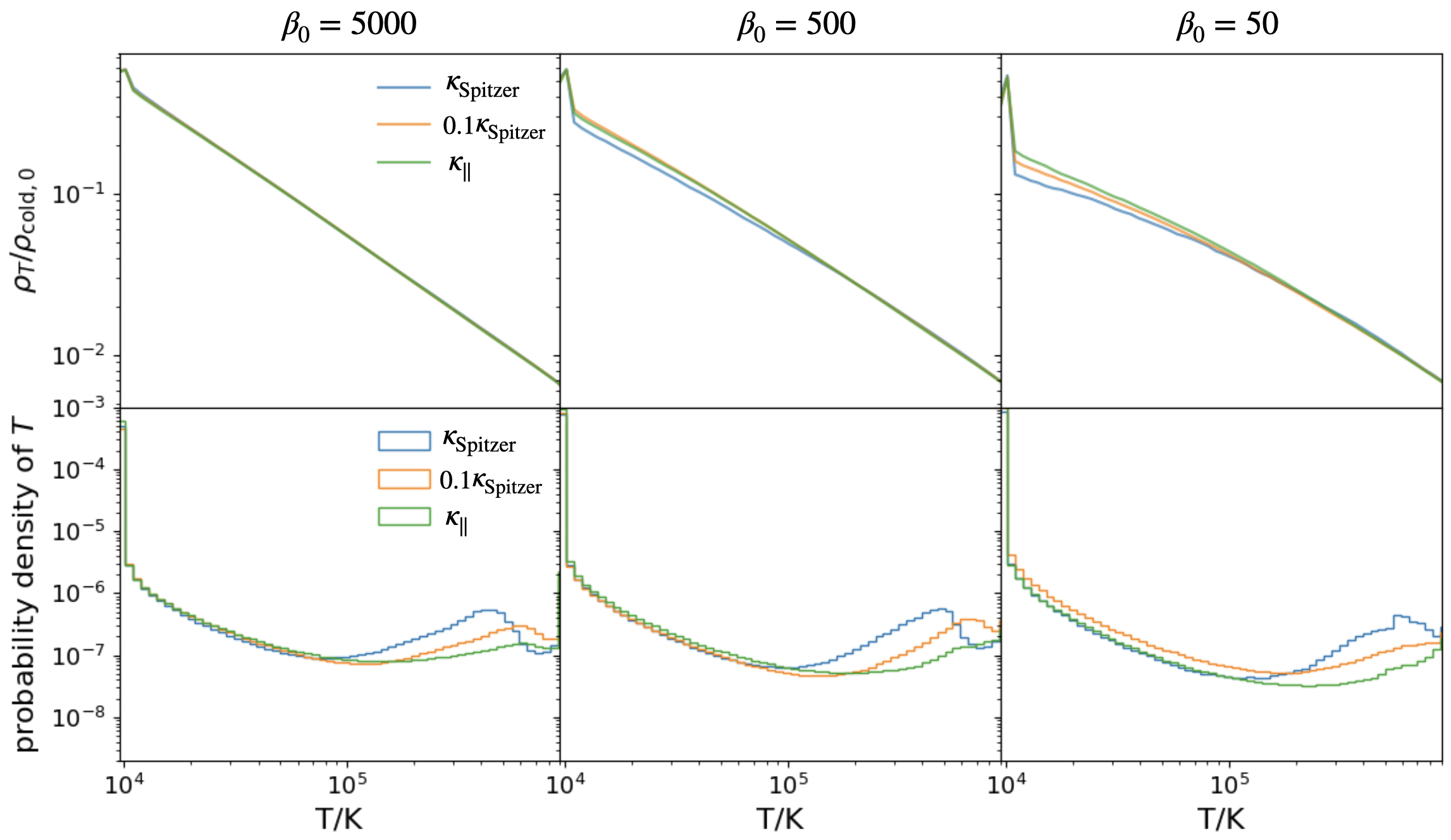}
\caption{\label{hist_Ksp}The mean density distributions as a function of temperature {\bf{(top)}} and the temperature PDFs {\bf{(bottom)}}, similar to Fig. \ref{hist_mag}. From the left to right columns, magnetization increases from $\beta_0=5000$ to $\beta_0=50$. Cooling is fiducial $(\Lambda_0=1)$ for all simulations here. The three colors represent different prescriptions of the conductivity coefficients. Regarding the temperature PDFs, there are only small differences in hot phases with minor contribution to the surface brightness $Q$.}

\end{center}
\end{figure*}

When adopting $\kappa_{\rm Spitzer}$ in hydrodynamic TMLs, \cite{Tan:21B} find temperature PDF is obviously shifted to higher temperatures and the cold phases are heavily suppressed compared with the simulations employing constant isotropic thermal conductivity. This is because $\kappa_{\rm Spitzer}$ is strongly temperature-dependent and substantially raises the conductivity in hot phase. In magnetized TMLs, we expect temperature PDFs to be less sensitive to the anisotropic conductivity since magnetic fields largely perpendicular to the overall temperature gradient should inhibit thermal conduction. Indeed, in Figure \ref{hist_Ksp}, we see different conductivites only make minor difference in the hot phases which modestly contribute to the surface brightness $Q$. The density distributions are barely affected, indicating $P_{\rm mag}$ is not sensitive to conductivity either.
Nevertheless, given the natural coupling of anisotropic conduction and magnetic field orientations, the role of conductivity should be further investigated with other field geometries.

\section{Discussion}\label{discussion}

In this section, we compare our work with previous results, and discuss potential implications to global scale simulations.

\subsection{Compare with previous TML works: scaling relations of $Q$}\label{frac_theory}
Main results of earlier works on hydrodynamic TMLs (e.g., \cite{Fielding:20,Tan:21}) can be roughly summarized by the two-piece scaling relation: $Q\propto t_{\rm cool}^{-1/2}$ and $Q\propto t_{\rm cool}^{-1/4}$, with the criterion ${\rm Da}=1$ governing the transition (equation \ref{Da_eq}). While the former half can be plainly elucidated by a laminar model, the latter half, however, requires more insight to understand since emerging fractal structure of the intermediate-temperature surface complicates the physics.


Borrowing wisdom from combustion theory, \cite{Tan:21} explained the $Q\propto t_{\rm cool}^{-1/4}$ scaling, whose success suggests similarities in the essential physics behind such two problems. On the other hand, \cite{Fielding:20} also derived the same scaling from a fractal point of view. They measured the surface of turbulent mixing front to have a fractal dimension $D=5/2$, and directly estimate $Q$ by definition (equation \ref{Q_def}) at the scale where cooling is the most efficient.

In spite of the good match between theories and hydrodynamic TML simulations, we see in Figure \ref{Q_scaling} they are no longer accurate even in relatively weakly magnetized environments ($\beta_0\lesssim 1000$). As has been discussed, magnetic fields can easily be amplified to reduce/suppress mixing, and can play a major in the pressure balance. Therefore, it fundamentally alters the balance between radiative cooling and turbulent mixing, and breaks the analogy with the combustion theory. On the other hand, the scaling predicted by the fractal theory is sensitive to the fractal dimension $D$, which is obtained by fitting the effective surface area $A_\lambda$ with fastest cooling rate as a function of scale $\lambda$, and $A_{\lambda}\propto \lambda^{-1/2}$ leads to $D=5/2$. Since the morphology of mixing layers is substantially different in MHD (Figure \ref{gallery} and \ref{section_mag}), we anticipate that the fractal nature changes as well. In fact, we have measured $D$ in our simulations following \cite{Fielding:20} and can approximately reproduce their result in pure hydrodynamic simulations without thermal conduction. However, we also find that the measurement could depend on the simulation setup, especially on the presence of thermal conduction, as well as the algorithm that extracts $A_\lambda$ \citep{Cintosun07}, which may not lead to a well-defined $D$ from fitting the $A_\lambda-\lambda$ relation at our resolution.
Nevertheless, conducting such measurement from our MHD simulations, we observe a clear flattening trend in the $A_\lambda-\lambda$ relation indicative of smaller $D$ (more laminar) towards higher magnetization. This is in line with our main findings.


\subsection{Links to global scale problems}

One main purpose for studying magnetized TMLs is to find its applications towards global scales, and our finding of heavily reduced $Q$ in mixing layers should imply inefficient energy transfer between cold/hot gas in magnetized environments. This is indeed observed by \cite{Gronnow18} in their cloud crushing simulations, where magnetic fields appreciably prevent the mass growth of cold clouds at early time. However, \cite{Gronke20a} reported that at later time, neither the cloud growth criterion nor the final mass growth is sensitive to magnetic fields, which appears inconsistent with our results. They raised two possibilities: (i) global simulations may not faithfully capture small scale interactions across TMLs. (ii) magnetic fields considerably change cloud morphology and increase the contact area between cold and hot gas, neutralizing suppressed local interactions. Given the convergence tests we have performed, and the fact that magnetic fields reshape the contact surface both locally (Figure \ref{section_mag}) and globally \citep{Banda-Barrag16,Cottle20,Gronke20a}, we expect the second explanation to be more likely to hold. Very recently, we notice \cite{KDas23} simultaneously studied both the magnetized TMLs and their connections to the growth of cold clouds. While they also found suppressed $Q$ in magnetized TMLs, they confirmed a lack of difference in cloud growth rates between MHD and hydrodynamic simulations. They attributed such paradox to that mixing is mostly governed by the intensity of turbulence: in TML simulations magnetic fields suppress turbulence by suppressing KHI, which can be overtaken by turbulence driven at large scales.
Besides there are uncertainties on the nature of large-scale turbulence, we also note that in simulations with external turbulence, the relative velocity between the clouds and background gas is vanishing, representing a different regime from typical TML simulations.
\cite{Li20} also ran MHD cloud-crushing simulations, reporting the role of magnetic fields being inconsequential. However, they acknowledge that the geometry of magnetic fields could make a difference, especially considering its coupling with thermal conduction.

Global scale MHD simulations typically embrace magnetic fields much stronger than our initial conditions \citep{McCourt15,Gronnow18,Cottle20}, so that magnetic fields can be dynamically important. However, there is yet no consensus answer for the realistic $\beta$ in the warm and hot CGM, while some estimations suggest $\beta\sim 10^2 - 10^9$ \citep{Su17,MartinA18,Hopkins20,Li20}. Our results therefore provide supplements to the weak field regime, and point out even quite weak magnetization can change the local dynamics and phase distribution, especially when radiative cooling is very efficient. For strongly magnetized TMLs, our plane-parallel model in limited simulation domain would be less appropriate, because strong magnetic fields can resist rolling by the KHI, and thus the overall dynamics is subject to the largely unknown global field configuration.
Under this situation, it has recently been suggested that the system can exhibit rich magnetothermal phase structures filled with plasmoids \citep{Fielding22}, a scenario that deserves further investigation in a global setting.

In addition, our results could provide corrections for a 1D effective model representing the contribution from TMLs. For example, \cite{Tan:21B} recently constructed such a hydrodynamic 1D model which enables sub-grid absorption and emission line predictions, and their predicted line ratios agree well with the observations. However in their model, assuming a Spitzer value conductivity would lead to significant deviations from the observations. This issue may potentially be resolved by taking into account of magnetic effects. As we discuss about Figure \ref{hist_Ksp}, even weak magnetic fields substantially prohibit the heat fluxes, and the temperature PDF only modestly shifts towards the hot phase when we use different conductivity prescriptions. The reduced $Q$ in magnetized TMLs also suggests that column densities should be lower than the predictions from 1D hydrodynamic models.

\section{Summary}\label{conclusion}

We have performed 3D MHD plane parallel simulations to examine the development of weakly magnetized TMLs, which is designed to mimic phase boundaries in a multiphase system (such as the CGM), aiming to compare and contrast existing studies of hydrodynamic TMLs. Imposing an initially uniform magnetic field along shear flow direction ($\mathbf{B_0}=B_0\hat{\mathbf{y}}$), we have found that even rather weakly magnetized environment $(\beta_0\sim 1000)$ can lead to substantial differences from the hydrodynamic cases. We list our main conclusions below:

\begin{enumerate}[wide, labelwidth=!,itemindent=!,labelindent=0pt, leftmargin=0em, label=(\arabic*), parsep=0pt]
    \item {\bf{Surface brightness $Q$}}
    
    The surface brightness $Q$, and hence the inflow velocities $v_{\rm in}$ for the growth of cold gas, can be substantially suppressed in magnetized TMLs even with a weak initial field (e.g., by a factor of $\sim10$ for $\beta_0=500$). There is a lack of specific scaling relations for $Q$ as a function of cooling strength $\Lambda$, in contrast to the hydrodynamic case, and
    the time evolution of $Q$ is more fluctuating.
    Within the range of fluctuations of $Q$, our results reach good convergence. The final level of $Q$ should be determined by the coupling of radiative cooling, turbulent mixing and magnetic field amplification.

    
    \item {\bf{Morphology of magnetized TML}}
    
    We observe weak initial fields $(\beta_0\geq 50)$ can be substantially amplified in the TML, resulting in highly magnetized $(\beta \sim 1)$ cold phase in the vicinity of the mixing layer. The intensified magnetic fields
    makes the mixing layer more ``fractal", and eventually more laminar (for stronger initial field and cooling),
    compared with hydrodynamic TMLs. The magnetic pressure $P_{\rm mag}$ has major contributions in the TML, so that the total pressure $P^*\equiv P_{\rm mag}+P_{\rm therm}$ is in good equilibrium across the magnetized TMLs.

    \item {\bf{Two distinct magnetic influences}}
    
    In magnetized TMLs, magnetic fields mainly reduce $Q$ in two ways: reducing gas pressure (and hence density and emissivity) by $P_{\rm mag}$, and directly suppressing turbulent mixing. 
    With very weak initial field $(\beta_0\gtrsim 5000)$ and/or weak cooling $(\Lambda_0\sim 0.1)$, $Q$ is mostly reduced by suppression of turbulent mixing. With stronger magnetic field and cooling, this suppression tends to saturate, and further reduction in $Q$ results from the TML being increasingly strongly magnetized with $P_{\rm mag}$ dominating over $P_{\rm therm}$.
    This could potentially explain the density imbalance between phases in the CGM, as inferred from observations \citep{Werk14}.
    
    \item {\bf{Initial field geometry}}

    We run simulations with initial magnetic fields $\mathbf{B}_0$ separately along three coordinate axes. When $\mathbf{B}_0$ is parallel to the cold/hot interface, the properties of magnetized TMLs are generally similar, in agreement with \cite{Ji:19}. Once $\mathbf{B}_0$ is normal to the interface, $Q$ is heavily suppressed due to continuous shear amplification of the initial field without establishing a pressure equilibrium.

    \item {\bf{Conductivity}}
    
    We compare three sets of simulations separately with constant conductivity $\kappa_{\parallel}$ (equation \ref{K_parallel}), Spitzer conductivity (equation \ref{K_Spitzer}) and reduced Spitzer conductivity $(0.1\kappa_{\rm Spitzer})$. Given the anisotropy in thermal conduction and our choice of field geometry, different choices of conductivity hardly change the result of $Q$, and only make minor differences in the temperature PDFs.


\end{enumerate}

    

Our simulations serve as an initial investigation to incorporate one important physics, namely, magnetic fields, on the dynamics of the multiphase ISM/CGM. While demonstrating its importance, the results are subject to a number of simplifications and caveats. The choice of our initial field geometry with uniform fields is somewhat artificial, and entangled magnetic field configuration deserves further investigations. Being local simulations with limited domain size, our study can be subject to artificial truncation of large-scale magnetic fields. This hinders us from exploring more strongly magnetized environments where global field geometry becomes more crucial. The fact that we find $Q$ is insensitive to resolution encourages migration towards cloud-scale simulations (e.g., \citealp{Gronke20a,Fielding:21,Tan23}). Also, in this work, we only examined one typical choice of density contrast $\chi\equiv \rho_{\rm cold}/\rho_{\rm hot}=100$ for cooling curves in typical CGM conditions. It remains to explore the parameter corresponding to other environments, such as $\chi\sim 4000$ in a ICM-like environment \citep{Qiu:20},
where hot gas entrainment into the TMLs is expected to be enhanced by larger $\chi$ \citep{Fielding:20}.

There are also additional missing physical ingredients. We did not consider the role of viscosity, though our preliminary investigation suggests it is unlikely to be important, as also found in \cite{Li20} in the context of the cloud crushing problem. A more important factor concerns the role of dynamically-important cosmic-rays (CRs), which can substantially alter the phase structure and energetics of the multiphase ISM/CGM  \cite[e.g.][]{Ji:20}. Given that the CRs and magnetic fields are inherently coupled, incorporating CRs represents another natural extension of our work towards better understanding the physics of the multiphase ISM/CGM. 
    
\section*{Acknowledgements}

We thank Suoqing Ji, Drummond Fielding, Eve Ostriker, Yu Qiu and Haitao Xu for helpful discussions and advice. XZ is also grateful to his friends for their unwavering encouragement. This research was supported by NSFC grant 11873033 and 12042505. Numerical simulations are conducted on TianHe-1 (A) at National Supercomputer Center in Tianjin, China, and on the Orion cluster at Department of Astronomy, Tsinghua University.

\section*{Data Availability}

 The data underlying this article will be shared on reasonable request to the corresponding author.



\bibliographystyle{mnras}
\bibliography{TMLBib} 

\begin{thebibliography}{}
\makeatletter
\relax
\def\mn@urlcharsother{\let\do\@makeother \do\$\do\&\do\#\do\^\do\_\do\%\do\~}
\def\mn@doi{\begingroup\mn@urlcharsother \@ifnextchar [ {\mn@doi@} {\mn@doi@[]}}
\def\mn@doi@[#1]#2{\def\@tempa{#1}\ifx\@tempa\@empty \href {http://dx.doi.org/#2} {doi:#2}\else \href {http://dx.doi.org/#2} {#1}\fi \endgroup}
\def\mn@eprint#1#2{\mn@eprint@#1:#2::\@nil}
\def\mn@eprint@arXiv#1{\href {http://arxiv.org/abs/#1} {{\tt arXiv:#1}}}
\def\mn@eprint@dblp#1{\href {http://dblp.uni-trier.de/rec/bibtex/#1.xml} {dblp:#1}}
\def\mn@eprint@#1:#2:#3:#4\@nil{\def\@tempa {#1}\def\@tempb {#2}\def\@tempc {#3}\ifx \@tempc \@empty \let \@tempc \@tempb \let \@tempb \@tempa \fi \ifx \@tempb \@empty \def\@tempb {arXiv}\fi \@ifundefined {mn@eprint@\@tempb}{\@tempb:\@tempc}{\expandafter \expandafter \csname mn@eprint@\@tempb\endcsname \expandafter{\@tempc}}}

\bibitem[\protect\citeauthoryear{{Armillotta}, {Fraternali}  \& {Marinacci}}{{Armillotta} et~al.}{2016}]{Armillotta16}
{Armillotta} L.,  {Fraternali} F.,   {Marinacci} F.,  2016, \mn@doi [\mnras] {10.1093/mnras/stw1930}, \href {https://ui.adsabs.harvard.edu/abs/2016MNRAS.462.4157A} {462, 4157}

\bibitem[\protect\citeauthoryear{{Banda-Barrag{\'a}n}, {Parkin}, {Federrath}, {Crocker}  \& {Bicknell}}{{Banda-Barrag{\'a}n} et~al.}{2016}]{Banda-Barrag16}
{Banda-Barrag{\'a}n} W.~E.,  {Parkin} E.~R.,  {Federrath} C.,  {Crocker} R.~M.,   {Bicknell} G.~V.,  2016, \mn@doi [\mnras] {10.1093/mnras/stv2405}, \href {https://ui.adsabs.harvard.edu/abs/2016MNRAS.455.1309B} {455, 1309}

\bibitem[\protect\citeauthoryear{{Banda-Barrag{\'a}n}, {Federrath}, {Crocker}  \& {Bicknell}}{{Banda-Barrag{\'a}n} et~al.}{2018a}]{BandaBarrag18}
{Banda-Barrag{\'a}n} W.~E.,  {Federrath} C.,  {Crocker} R.~M.,   {Bicknell} G.~V.,  2018a, \mn@doi [\mnras] {10.1093/mnras/stx2541}, \href {https://ui.adsabs.harvard.edu/abs/2018MNRAS.473.3454B} {473, 3454}

\bibitem[\protect\citeauthoryear{{Banda-Barrag{\'a}n}, {Federrath}, {Crocker}  \& {Bicknell}}{{Banda-Barrag{\'a}n} et~al.}{2018b}]{BB17}
{Banda-Barrag{\'a}n} W.~E.,  {Federrath} C.,  {Crocker} R.~M.,   {Bicknell} G.~V.,  2018b, \mn@doi [\mnras] {10.1093/mnras/stx2541}, \href {https://ui.adsabs.harvard.edu/abs/2018MNRAS.473.3454B} {473, 3454}

\bibitem[\protect\citeauthoryear{{Begelman} \& {Fabian}}{{Begelman} \& {Fabian}}{1990}]{Begelman90}
{Begelman} M.~C.,  {Fabian} A.~C.,  1990, \mnras, \href {https://ui.adsabs.harvard.edu/abs/1990MNRAS.244P..26B} {244, 26P}

\bibitem[\protect\citeauthoryear{{Begelman} \& {McKee}}{{Begelman} \& {McKee}}{1990}]{Begelman&McKee90}
{Begelman} M.~C.,  {McKee} C.~F.,  1990, \mn@doi [\apj] {10.1086/168994}, \href {https://ui.adsabs.harvard.edu/abs/1990ApJ...358..375B} {358, 375}

\bibitem[\protect\citeauthoryear{{Borkowski}, {Balbus}  \& {Fristrom}}{{Borkowski} et~al.}{1990}]{Borkowski:90}
{Borkowski} K.~J.,  {Balbus} S.~A.,   {Fristrom} C.~C.,  1990, \mn@doi [\apj] {10.1086/168784}, \href {https://ui.adsabs.harvard.edu/abs/1990ApJ...355..501B/abstract} {355, 501}

\bibitem[\protect\citeauthoryear{{Brandenburg} \& {Subramanian}}{{Brandenburg} \& {Subramanian}}{2005}]{Brandenburg05}
{Brandenburg} A.,  {Subramanian} K.,  2005, \mn@doi [\physrep] {10.1016/j.physrep.2005.06.005}, \href {https://ui.adsabs.harvard.edu/abs/2005PhR...417....1B} {417, 1}

\bibitem[\protect\citeauthoryear{{Braspenning}, {Schaye}, {Borrow}  \& {Schaller}}{{Braspenning} et~al.}{2022}]{Braspenning22}
{Braspenning} J.,  {Schaye} J.,  {Borrow} J.,   {Schaller} M.,  2022, arXiv e-prints, \href {https://ui.adsabs.harvard.edu/abs/2022arXiv220313915B} {p. arXiv:2203.13915}

\bibitem[\protect\citeauthoryear{{Br{\"u}ggen} \& {Scannapieco}}{{Br{\"u}ggen} \& {Scannapieco}}{2016}]{Bruggen16}
{Br{\"u}ggen} M.,  {Scannapieco} E.,  2016, \mn@doi [\apj] {10.3847/0004-637X/822/1/31}, \href {https://ui.adsabs.harvard.edu/abs/2016ApJ...822...31B} {822, 31}

\bibitem[\protect\citeauthoryear{{Br{\"u}ggen} \& {Scannapieco}}{{Br{\"u}ggen} \& {Scannapieco}}{2020}]{Bruggen20}
{Br{\"u}ggen} M.,  {Scannapieco} E.,  2020, \mn@doi [\apj] {10.3847/1538-4357/abc00f}, \href {https://ui.adsabs.harvard.edu/abs/2020ApJ...905...19B} {905, 19}

\bibitem[\protect\citeauthoryear{{Br{\"u}ggen}, {Scannapieco}  \& {Grete}}{{Br{\"u}ggen} et~al.}{2023}]{Bruggen23}
{Br{\"u}ggen} M.,  {Scannapieco} E.,   {Grete} P.,  2023, \mn@doi [arXiv e-prints] {10.48550/arXiv.2304.09881}, \href {https://ui.adsabs.harvard.edu/abs/2023arXiv230409881B} {p. arXiv:2304.09881}

\bibitem[\protect\citeauthoryear{{Bustard} \& {Gronke}}{{Bustard} \& {Gronke}}{2022}]{Bustard22}
{Bustard} C.,  {Gronke} M.,  2022, \mn@doi [\apj] {10.3847/1538-4357/ac752b}, \href {https://ui.adsabs.harvard.edu/abs/2022ApJ...933..120B} {933, 120}

\bibitem[\protect\citeauthoryear{{Chandrasekar}}{{Chandrasekar}}{1961}]{Chandra:61}
{Chandrasekar} S.,  1961, {Hydrodynamic and Hydromagnetic Stability}.
Dover Publications, INC. New York

\bibitem[\protect\citeauthoryear{{Chen} \& {Mulchaey}}{{Chen} \& {Mulchaey}}{2009}]{Chen09}
{Chen} H.-W.,  {Mulchaey} J.~S.,  2009, \mn@doi [\apj] {10.1088/0004-637X/701/2/1219}, \href {https://ui.adsabs.harvard.edu/abs/2009ApJ...701.1219C} {701, 1219}

\bibitem[\protect\citeauthoryear{{Chen}, {Helsby}, {Gauthier}, {Shectman}, {Thompson}  \& {Tinker}}{{Chen} et~al.}{2010}]{Chen10}
{Chen} H.-W.,  {Helsby} J.~E.,  {Gauthier} J.-R.,  {Shectman} S.~A.,  {Thompson} I.~B.,   {Tinker} J.~L.,  2010, \mn@doi [\apj] {10.1088/0004-637X/714/2/1521}, \href {https://ui.adsabs.harvard.edu/abs/2010ApJ...714.1521C} {714, 1521}

\bibitem[\protect\citeauthoryear{{Cintosun}, {Smallwood}  \& {G{\"u}lder}}{{Cintosun} et~al.}{2007}]{Cintosun07}
{Cintosun} E.,  {Smallwood} G.~J.,   {G{\"u}lder} {\"O}.~L.,  2007, \mn@doi [AIAA Journal] {10.2514/1.29533}, \href {https://ui.adsabs.harvard.edu/abs/2007AIAAJ..45.2785C} {45, 2785}

\bibitem[\protect\citeauthoryear{{Cottle}, {Scannapieco}, {Br{\"u}ggen}, {Banda-Barrag{\'a}n}  \& {Federrath}}{{Cottle} et~al.}{2020}]{Cottle20}
{Cottle} J.,  {Scannapieco} E.,  {Br{\"u}ggen} M.,  {Banda-Barrag{\'a}n} W.,   {Federrath} C.,  2020, \mn@doi [\apj] {10.3847/1538-4357/ab76d1}, \href {https://ui.adsabs.harvard.edu/abs/2020ApJ...892...59C} {892, 59}

\bibitem[\protect\citeauthoryear{{Cowie} \& {McKee}}{{Cowie} \& {McKee}}{1977}]{Cowie77}
{Cowie} L.~L.,  {McKee} C.~F.,  1977, \mn@doi [\apj] {10.1086/154911}, \href {https://ui.adsabs.harvard.edu/abs/1977ApJ...211..135C} {211, 135}

\bibitem[\protect\citeauthoryear{{Damköhler}}{{Damköhler}}{1940}]{Damk:40}
{Damköhler} G.,  1940, \mn@doi [Zeitschrift für Elektrochemie und angewandte physikalische Chemie] {10.1002/bbpc.19400461102}, \href {https://onlinelibrary.wiley.com/doi/10.1002/bbpc.19400461102} {46, 601}

\bibitem[\protect\citeauthoryear{{Das} \& {Gronke}}{{Das} \& {Gronke}}{2023}]{KDas23}
{Das} H.~K.,  {Gronke} M.,  2023, \mn@doi [arXiv e-prints] {10.48550/arXiv.2307.06411}, \href {https://ui.adsabs.harvard.edu/abs/2023arXiv230706411D} {p. arXiv:2307.06411}

\bibitem[\protect\citeauthoryear{{Drake} et~al.,}{{Drake} et~al.}{2021}]{Drake21}
{Drake} J.~F.,  et~al., 2021, \mn@doi [\apj] {10.3847/1538-4357/ac1ff1}, \href {https://ui.adsabs.harvard.edu/abs/2021ApJ...923..245D} {923, 245}

\bibitem[\protect\citeauthoryear{{Dursi} \& {Pfrommer}}{{Dursi} \& {Pfrommer}}{2008}]{Dursi08}
{Dursi} L.~J.,  {Pfrommer} C.,  2008, \mn@doi [\apj] {10.1086/529371}, \href {https://ui.adsabs.harvard.edu/abs/2008ApJ...677..993D} {677, 993}

\bibitem[\protect\citeauthoryear{{El-Badry}, {Ostriker}, {Kim}, {Quataert}  \& {Weisz}}{{El-Badry} et~al.}{2019}]{ElBadry19}
{El-Badry} K.,  {Ostriker} E.~C.,  {Kim} C.-G.,  {Quataert} E.,   {Weisz} D.~R.,  2019, \mn@doi [\mnras] {10.1093/mnras/stz2773}, \href {https://ui.adsabs.harvard.edu/abs/2019MNRAS.490.1961E} {490, 1961}

\bibitem[\protect\citeauthoryear{{Faucher-Giguere} \& {Oh}}{{Faucher-Giguere} \& {Oh}}{2023}]{FGO23}
{Faucher-Giguere} C.-A.,  {Oh} S.~P.,  2023, \mn@doi [arXiv e-prints] {10.48550/arXiv.2301.10253}, \href {https://ui.adsabs.harvard.edu/abs/2023arXiv230110253F} {p. arXiv:2301.10253}

\bibitem[\protect\citeauthoryear{{Federrath}}{{Federrath}}{2016}]{Federrath16}
{Federrath} C.,  2016, \mn@doi [Journal of Plasma Physics] {10.1017/S0022377816001069}, \href {https://ui.adsabs.harvard.edu/abs/2016JPlPh..82f5301F} {82, 535820601}

\bibitem[\protect\citeauthoryear{{Fielding} \& {Bryan}}{{Fielding} \& {Bryan}}{2022}]{Fielding:21}
{Fielding} D.~B.,  {Bryan} G.~L.,  2022, \mn@doi [\apj] {10.3847/1538-4357/ac2f41}, \href {https://ui.adsabs.harvard.edu/abs/2022ApJ...924...82F} {924, 82}

\bibitem[\protect\citeauthoryear{{Fielding}, {Quataert}  \& {Martizzi}}{{Fielding} et~al.}{2018}]{Fielding18}
{Fielding} D.,  {Quataert} E.,   {Martizzi} D.,  2018, \mn@doi [\mnras] {10.1093/mnras/sty2466}, \href {https://ui.adsabs.harvard.edu/abs/2018MNRAS.481.3325F} {481, 3325}

\bibitem[\protect\citeauthoryear{{Fielding}, {Ostriker}, {Bryan}  \& {Jermyn}}{{Fielding} et~al.}{2020}]{Fielding:20}
{Fielding} D.~B.,  {Ostriker} E.~C.,  {Bryan} G.~L.,   {Jermyn} A.~S.,  2020, \mn@doi [\apjl] {10.3847/2041-8213/ab8d2c}, \href {https://ui.adsabs.harvard.edu/abs/2020ApJ...894L..24F} {894, L24}

\bibitem[\protect\citeauthoryear{{Fielding}, {Ripperda}  \& {Philippov}}{{Fielding} et~al.}{2022}]{Fielding22}
{Fielding} D.~B.,  {Ripperda} B.,   {Philippov} A.~A.,  2022, \mn@doi [arXiv e-prints] {10.48550/arXiv.2211.06434}, \href {https://ui.adsabs.harvard.edu/abs/2022arXiv221106434F} {p. arXiv:2211.06434}

\bibitem[\protect\citeauthoryear{{Gnat} \& {Sternberg}}{{Gnat} \& {Sternberg}}{2007}]{Gnat:07}
{Gnat} O.,  {Sternberg} A.,  2007, \mn@doi [\apjs] {10.1086/509786}, \href {https://ui.adsabs.harvard.edu/abs/2007ApJS..168..213G} {168, 213}

\bibitem[\protect\citeauthoryear{{Gnat}, {Sternberg}  \& {McKee}}{{Gnat} et~al.}{2010}]{Gnat:10}
{Gnat} O.,  {Sternberg} A.,   {McKee} C.~F.,  2010, \apj, \href {https://iopscience.iop.org/article/10.1088/0004-637X/718/2/1315} {718, 1315}

\bibitem[\protect\citeauthoryear{{Gronke} \& {Oh}}{{Gronke} \& {Oh}}{2018}]{Gronke:18}
{Gronke} M.,  {Oh} S.~P.,  2018, \mn@doi [\mnras] {10.1093/mnrasl/sly131}, \href {https://academic.oup.com/mnrasl/article/480/1/L111/5056723} {480, L111}

\bibitem[\protect\citeauthoryear{{Gronke} \& {Oh}}{{Gronke} \& {Oh}}{2020}]{Gronke20a}
{Gronke} M.,  {Oh} S.~P.,  2020, \mn@doi [\mnras] {10.1093/mnras/stz3332}, \href {https://ui.adsabs.harvard.edu/abs/2020MNRAS.492.1970G} {492, 1970}

\bibitem[\protect\citeauthoryear{{Gr{\o}nnow}, {Tepper-Garc{\'\i}a}  \& {Bland-Hawthorn}}{{Gr{\o}nnow} et~al.}{2018}]{Gronnow18}
{Gr{\o}nnow} A.,  {Tepper-Garc{\'\i}a} T.,   {Bland-Hawthorn} J.,  2018, \mn@doi [\apj] {10.3847/1538-4357/aada0e}, \href {https://ui.adsabs.harvard.edu/abs/2018ApJ...865...64G} {865, 64}

\bibitem[\protect\citeauthoryear{{Hennawi}, {Prochaska}, {Cantalupo}  \& {Arrigoni-Battaia}}{{Hennawi} et~al.}{2015}]{Hennawi15}
{Hennawi} J.~F.,  {Prochaska} J.~X.,  {Cantalupo} S.,   {Arrigoni-Battaia} F.,  2015, \mn@doi [Science] {10.1126/science.aaa5397}, \href {https://ui.adsabs.harvard.edu/abs/2015Sci...348..779H} {348, 779}

\bibitem[\protect\citeauthoryear{{Hopkins} et~al.,}{{Hopkins} et~al.}{2020}]{Hopkins20}
{Hopkins} P.~F.,  et~al., 2020, \mn@doi [\mnras] {10.1093/mnras/stz3321}, \href {https://ui.adsabs.harvard.edu/abs/2020MNRAS.492.3465H} {492, 3465}

\bibitem[\protect\citeauthoryear{{Ji}, {Oh}  \& {Masterson}}{{Ji} et~al.}{2019}]{Ji:19}
{Ji} S.,  {Oh} S.~P.,   {Masterson} P.,  2019, \mn@doi [\mnras] {10.1093/mnras/stz1248}, \href {https://ui.adsabs.harvard.edu/abs/2019MNRAS.487..737J} {487, 737}

\bibitem[\protect\citeauthoryear{{Ji} et~al.,}{{Ji} et~al.}{2020}]{Ji:20}
{Ji} S.,  et~al., 2020, \mn@doi [\mnras] {10.1093/mnras/staa1849}, \href {https://ui.adsabs.harvard.edu/abs/2020MNRAS.496.4221J} {496, 4221}

\bibitem[\protect\citeauthoryear{{Jones}, {Gaalaas}, {Ryu}  \& {Frank}}{{Jones} et~al.}{1997}]{Jones:97}
{Jones} T.~W.,  {Gaalaas} J.~B.,  {Ryu} D.,   {Frank} A.,  1997, \mn@doi [\apj] {10.1086/304145}, \href {https://ui.adsabs.harvard.edu/abs/1997ApJ...482..230J} {482, 230}

\bibitem[\protect\citeauthoryear{{Kim}, {Ostriker}  \& {Raileanu}}{{Kim} et~al.}{2017}]{Kim17}
{Kim} C.-G.,  {Ostriker} E.~C.,   {Raileanu} R.,  2017, \mn@doi [\apj] {10.3847/1538-4357/834/1/25}, \href {https://ui.adsabs.harvard.edu/abs/2017ApJ...834...25K} {834, 25}

\bibitem[\protect\citeauthoryear{{Klein}, {McKee}  \& {Colella}}{{Klein} et~al.}{1994}]{Klein94}
{Klein} R.~I.,  {McKee} C.~F.,   {Colella} P.,  1994, \mn@doi [\apj] {10.1086/173554}, \href {https://ui.adsabs.harvard.edu/abs/1994ApJ...420..213K} {420, 213}

\bibitem[\protect\citeauthoryear{{Komarov}, {Schekochihin}, {Churazov}  \& {Spitkovsky}}{{Komarov} et~al.}{2018}]{Komarov18}
{Komarov} S.,  {Schekochihin} A.~A.,  {Churazov} E.,   {Spitkovsky} A.,  2018, \mn@doi [Journal of Plasma Physics] {10.1017/S0022377818000399}, \href {https://ui.adsabs.harvard.edu/abs/2018JPlPh..84c9005K} {84, 905840305}

\bibitem[\protect\citeauthoryear{{Lancaster}, {Ostriker}, {Kim}  \& {Kim}}{{Lancaster} et~al.}{2021}]{Lancaster21}
{Lancaster} L.,  {Ostriker} E.~C.,  {Kim} J.-G.,   {Kim} C.-G.,  2021, \mn@doi [\apj] {10.3847/1538-4357/abf8ab}, \href {https://ui.adsabs.harvard.edu/abs/2021ApJ...914...89L} {914, 89}

\bibitem[\protect\citeauthoryear{{Li}, {Hopkins}, {Squire}  \& {Hummels}}{{Li} et~al.}{2020}]{Li20}
{Li} Z.,  {Hopkins} P.~F.,  {Squire} J.,   {Hummels} C.,  2020, \mn@doi [\mnras] {10.1093/mnras/stz3567}, \href {https://ui.adsabs.harvard.edu/abs/2020MNRAS.492.1841L} {492, 1841}

\bibitem[\protect\citeauthoryear{{Mandelker}, {Nagai}, {Aung}, {Dekel}, {Birnboim}  \& {van den Bosch}}{{Mandelker} et~al.}{2020}]{Mandelker20}
{Mandelker} N.,  {Nagai} D.,  {Aung} H.,  {Dekel} A.,  {Birnboim} Y.,   {van den Bosch} F.~C.,  2020, \mn@doi [\mnras] {10.1093/mnras/staa812}, \href {https://ui.adsabs.harvard.edu/abs/2020MNRAS.494.2641M} {494, 2641}

\bibitem[\protect\citeauthoryear{{Martin-Alvarez}, {Devriendt}, {Slyz}  \& {Teyssier}}{{Martin-Alvarez} et~al.}{2018}]{MartinA18}
{Martin-Alvarez} S.,  {Devriendt} J.,  {Slyz} A.,   {Teyssier} R.,  2018, \mn@doi [\mnras] {10.1093/mnras/sty1623}, \href {https://ui.adsabs.harvard.edu/abs/2018MNRAS.479.3343M} {479, 3343}

\bibitem[\protect\citeauthoryear{{McCourt}, {O'Leary}, {Madigan}  \& {Quataert}}{{McCourt} et~al.}{2015}]{McCourt15}
{McCourt} M.,  {O'Leary} R.~M.,  {Madigan} A.-M.,   {Quataert} E.,  2015, \mn@doi [\mnras] {10.1093/mnras/stv355}, \href {https://ui.adsabs.harvard.edu/abs/2015MNRAS.449....2M} {449, 2}

\bibitem[\protect\citeauthoryear{{Meinecke} et~al.,}{{Meinecke} et~al.}{2022}]{Meinecke22}
{Meinecke} J.,  et~al., 2022, \mn@doi [Science Advances] {10.1126/sciadv.abj6799}, \href {https://ui.adsabs.harvard.edu/abs/2022SciA....8J6799M} {8, eabj6799}

\bibitem[\protect\citeauthoryear{{Miura} \& {Pritchett}}{{Miura} \& {Pritchett}}{1982}]{Miura:82}
{Miura} A.,  {Pritchett} P.~L.,  1982, \mn@doi [Journal of Geophysical Research] {10.1029/JA087iA09p07431}, 87, 7431

\bibitem[\protect\citeauthoryear{{Prochaska}, {Weiner}, {Chen}, {Mulchaey}  \& {Cooksey}}{{Prochaska} et~al.}{2011}]{Prochaska11a}
{Prochaska} J.~X.,  {Weiner} B.,  {Chen} H.~W.,  {Mulchaey} J.,   {Cooksey} K.,  2011, \mn@doi [\apj] {10.1088/0004-637X/740/2/91}, \href {https://ui.adsabs.harvard.edu/abs/2011ApJ...740...91P} {740, 91}

\bibitem[\protect\citeauthoryear{{Prochaska} et~al.,}{{Prochaska} et~al.}{2017}]{Prochaska17}
{Prochaska} J.~X.,  et~al., 2017, \mn@doi [\apj] {10.3847/1538-4357/aa6007}, \href {https://ui.adsabs.harvard.edu/abs/2017ApJ...837..169P} {837, 169}

\bibitem[\protect\citeauthoryear{{Qiu}, {Bogdanovi{\'c}}, {Li}, {McDonald}  \& {McNamara}}{{Qiu} et~al.}{2020}]{Qiu:20}
{Qiu} Y.,  {Bogdanovi{\'c}} T.,  {Li} Y.,  {McDonald} M.,   {McNamara} B.~R.,  2020, \mn@doi [Nature Astronomy] {10.1038/s41550-020-1090-7}, \href {https://ui.adsabs.harvard.edu/abs/2020NatAs...4..900Q} {4, 900}

\bibitem[\protect\citeauthoryear{{Roberg-Clark}, {Drake}, {Reynolds}  \& {Swisdak}}{{Roberg-Clark} et~al.}{2016}]{RC16}
{Roberg-Clark} G.~T.,  {Drake} J.~F.,  {Reynolds} C.~S.,   {Swisdak} M.,  2016, \mn@doi [\apjl] {10.3847/2041-8205/830/1/L9}, \href {https://ui.adsabs.harvard.edu/abs/2016ApJ...830L...9R} {830, L9}

\bibitem[\protect\citeauthoryear{{Ryu}, {Jones}  \& {Frank}}{{Ryu} et~al.}{2000}]{Ryu:20}
{Ryu} D.,  {Jones} T.~W.,   {Frank} A.,  2000, \mn@doi [\apj] {10.1086/317789}, \href {https://ui.adsabs.harvard.edu/abs/2000ApJ...545..475R} {545, 475}

\bibitem[\protect\citeauthoryear{{Savage}, {Kim}, {Wakker}, {Keeney}, {Shull}, {Stocke}  \& {Green}}{{Savage} et~al.}{2014}]{Savage14}
{Savage} B.~D.,  {Kim} T.~S.,  {Wakker} B.~P.,  {Keeney} B.,  {Shull} J.~M.,  {Stocke} J.~T.,   {Green} J.~C.,  2014, \mn@doi [\apjs] {10.1088/0067-0049/212/1/8}, \href {https://ui.adsabs.harvard.edu/abs/2014ApJS..212....8S} {212, 8}

\bibitem[\protect\citeauthoryear{{Scannapieco} \& {Br{\"u}ggen}}{{Scannapieco} \& {Br{\"u}ggen}}{2015}]{Scannapieco15}
{Scannapieco} E.,  {Br{\"u}ggen} M.,  2015, \mn@doi [\apj] {10.1088/0004-637X/805/2/158}, \href {https://ui.adsabs.harvard.edu/abs/2015ApJ...805..158S} {805, 158}

\bibitem[\protect\citeauthoryear{{Sharma} \& {Hammett}}{{Sharma} \& {Hammett}}{2007}]{Sharma07}
{Sharma} P.,  {Hammett} G.~W.,  2007, \mn@doi [Journal of Computational Physics] {10.1016/j.jcp.2007.07.026}, \href {https://ui.adsabs.harvard.edu/abs/2007JCoPh.227..123S} {227, 123}

\bibitem[\protect\citeauthoryear{{Spitzer}}{{Spitzer}}{1962}]{Spitzer:62}
{Spitzer} L.,  1962, {Physics of fully ionized gases}.
Interscience Publishers, New York

\bibitem[\protect\citeauthoryear{{Stern}, {Hennawi}, {Prochaska}  \& {Werk}}{{Stern} et~al.}{2016}]{Stern16}
{Stern} J.,  {Hennawi} J.~F.,  {Prochaska} J.~X.,   {Werk} J.~K.,  2016, \mn@doi [\apj] {10.3847/0004-637X/830/2/87}, \href {https://ui.adsabs.harvard.edu/abs/2016ApJ...830...87S} {830, 87}

\bibitem[\protect\citeauthoryear{{Stocke}, {Keeney}, {Danforth}, {Shull}, {Froning}, {Green}, {Penton}  \& {Savage}}{{Stocke} et~al.}{2013}]{Stocke13}
{Stocke} J.~T.,  {Keeney} B.~A.,  {Danforth} C.~W.,  {Shull} J.~M.,  {Froning} C.~S.,  {Green} J.~C.,  {Penton} S.~V.,   {Savage} B.~D.,  2013, \mn@doi [\apj] {10.1088/0004-637X/763/2/148}, \href {https://ui.adsabs.harvard.edu/abs/2013ApJ...763..148S} {763, 148}

\bibitem[\protect\citeauthoryear{{Stone}, {Tomida}, {White}  \& {Felker}}{{Stone} et~al.}{2020}]{Athena:20}
{Stone} J.~M.,  {Tomida} K.,  {White} C.~J.,   {Felker} K.~G.,  2020, \mn@doi [\apjs] {10.3847/1538-4365/ab929b}, \href {https://ui.adsabs.harvard.edu/abs/2020ApJS..249....4S} {249, 4}

\bibitem[\protect\citeauthoryear{{Su}, {Hopkins}, {Hayward}, {Faucher-Gigu{\`e}re}, {Kere{\v{s}}}, {Ma}  \& {Robles}}{{Su} et~al.}{2017}]{Su17}
{Su} K.-Y.,  {Hopkins} P.~F.,  {Hayward} C.~C.,  {Faucher-Gigu{\`e}re} C.-A.,  {Kere{\v{s}}} D.,  {Ma} X.,   {Robles} V.~H.,  2017, \mn@doi [\mnras] {10.1093/mnras/stx1463}, \href {https://ui.adsabs.harvard.edu/abs/2017MNRAS.471..144S} {471, 144}

\bibitem[\protect\citeauthoryear{{Tan} \& {Fielding}}{{Tan} \& {Fielding}}{2023}]{Tan23}
{Tan} B.,  {Fielding} D.~B.,  2023, \mn@doi [arXiv e-prints] {10.48550/arXiv.2305.14424}, \href {https://ui.adsabs.harvard.edu/abs/2023arXiv230514424T} {p. arXiv:2305.14424}

\bibitem[\protect\citeauthoryear{{Tan} \& {Oh}}{{Tan} \& {Oh}}{2021}]{Tan:21B}
{Tan} B.,  {Oh} S.~P.,  2021, \mn@doi [\mnras] {10.1093/mnrasl/slab100}, \href {https://ui.adsabs.harvard.edu/abs/2021MNRAS.508L..37T/abstract} {508, L37}

\bibitem[\protect\citeauthoryear{{Tan}, {Oh}  \& {Gronke}}{{Tan} et~al.}{2021}]{Tan:21}
{Tan} B.,  {Oh} S.~P.,   {Gronke} M.,  2021, \mn@doi [\mnras] {10.1093/mnras/stab053}, \href {https://ui.adsabs.harvard.edu/abs/2021MNRAS.502.3179T} {502, 3179}

\bibitem[\protect\citeauthoryear{{Tan}, {Oh}  \& {Gronke}}{{Tan} et~al.}{2022}]{Tan22}
{Tan} B.,  {Oh} S.~P.,   {Gronke} M.,  2022, arXiv e-prints, \href {https://ui.adsabs.harvard.edu/abs/2022arXiv221006493T} {p. arXiv:2210.06493}

\bibitem[\protect\citeauthoryear{{Tumlinson}, {Peeples}, {Werk}  \& K.}{{Tumlinson} et~al.}{2017}]{Tumlinson:17}
{Tumlinson} J.,  {Peeples} M.~S.,  {Werk}  K. J.,  2017, \mn@doi [\araa] {10.1146/annurev-astro-091916-055240}, \href {https://ui.adsabs.harvard.edu/abs/2017ARA\%26A..55..389T/abstract} {55, 389}

\bibitem[\protect\citeauthoryear{{Werk}, {Prochaska}, {Thom}, {Tumlinson}, {Tripp}, {O'Meara}  \& {Meiring}}{{Werk} et~al.}{2012}]{Werk12}
{Werk} J.~K.,  {Prochaska} J.~X.,  {Thom} C.,  {Tumlinson} J.,  {Tripp} T.~M.,  {O'Meara} J.~M.,   {Meiring} J.~D.,  2012, \mn@doi [\apjs] {10.1088/0067-0049/198/1/3}, \href {https://ui.adsabs.harvard.edu/abs/2012ApJS..198....3W} {198, 3}

\bibitem[\protect\citeauthoryear{{Werk} et~al.,}{{Werk} et~al.}{2014}]{Werk14}
{Werk} J.~K.,  et~al., 2014, \mn@doi [\apj] {10.1088/0004-637X/792/1/8}, \href {https://ui.adsabs.harvard.edu/abs/2014ApJ...792....8W} {792, 8}

\bibitem[\protect\citeauthoryear{{Wiener}, {Zweibel}  \& {Ruszkowski}}{{Wiener} et~al.}{2019}]{Wiener19}
{Wiener} J.,  {Zweibel} E.~G.,   {Ruszkowski} M.,  2019, \mn@doi [\mnras] {10.1093/mnras/stz2007}, \href {https://ui.adsabs.harvard.edu/abs/2019MNRAS.489..205W} {489, 205}

\bibitem[\protect\citeauthoryear{{Xu} \& {Stone}}{{Xu} \& {Stone}}{1995}]{Xu95}
{Xu} J.,  {Stone} J.~M.,  1995, \mn@doi [\apj] {10.1086/176475}, \href {https://ui.adsabs.harvard.edu/abs/1995ApJ...454..172X} {454, 172}

\bibitem[\protect\citeauthoryear{{Yang} \& {Ji}}{{Yang} \& {Ji}}{2023}]{Yang23}
{Yang} Y.,  {Ji} S.,  2023, \mn@doi [\mnras] {10.1093/mnras/stad264}, \href {https://ui.adsabs.harvard.edu/abs/2023MNRAS.520.2148Y} {520, 2148}

\makeatother
\end{thebibliography}




\appendix

\section{Convergence Tests}\label{Appendix_conv}

While in Figure \ref{Q_conv} we have shown convergence study on $Q$ to verify our results, here we provide supplementary convergence tests on other aspects of our simulations.

\begin{figure*}
\begin{center}
\includegraphics[scale=0.41]{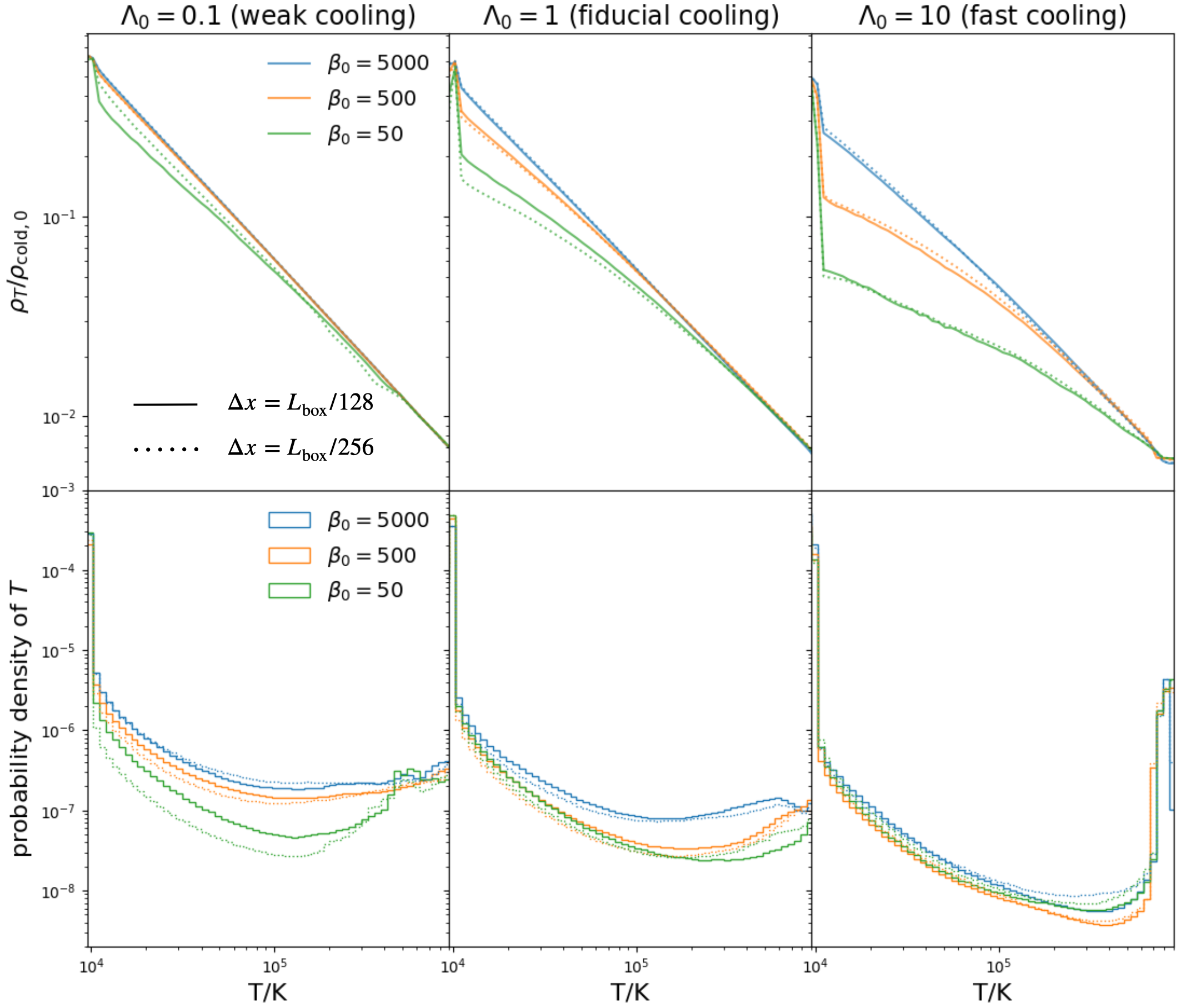}
\caption{\label{hist_conv}Convergence test on the mean density as a function of temperature {\bf{(top)}} and temperature PDFs {\bf{(bottom)}}. Solid lines are from simulations with fiducial resolution $(\Delta x=L_{\rm box}/128)$, while dotted lines represent doubled resolution $(\Delta x=L_{\rm box}/256)$.}

\end{center}
\end{figure*}

Figure \ref{hist_conv} displays the density distributions and temperature PDFs in simulations with fiducial resolution $(\Delta x=L_{\rm box}/128,{\ \rm solid\ lines})$ and doubled resolution $(\Delta x=L_{\rm box}/256,{\ \rm dotted\ lines})$. Generally speaking, convergence is good. Although for $\beta_0=50$ cases there can be slight deviations at the hot end of the temperature PDFs, these hot phases have little contribution to the surface brightness $Q$ and should be inconsequential especially considering the level of fluctuations.

\begin{figure*}
\begin{center}

\includegraphics[scale=0.41]{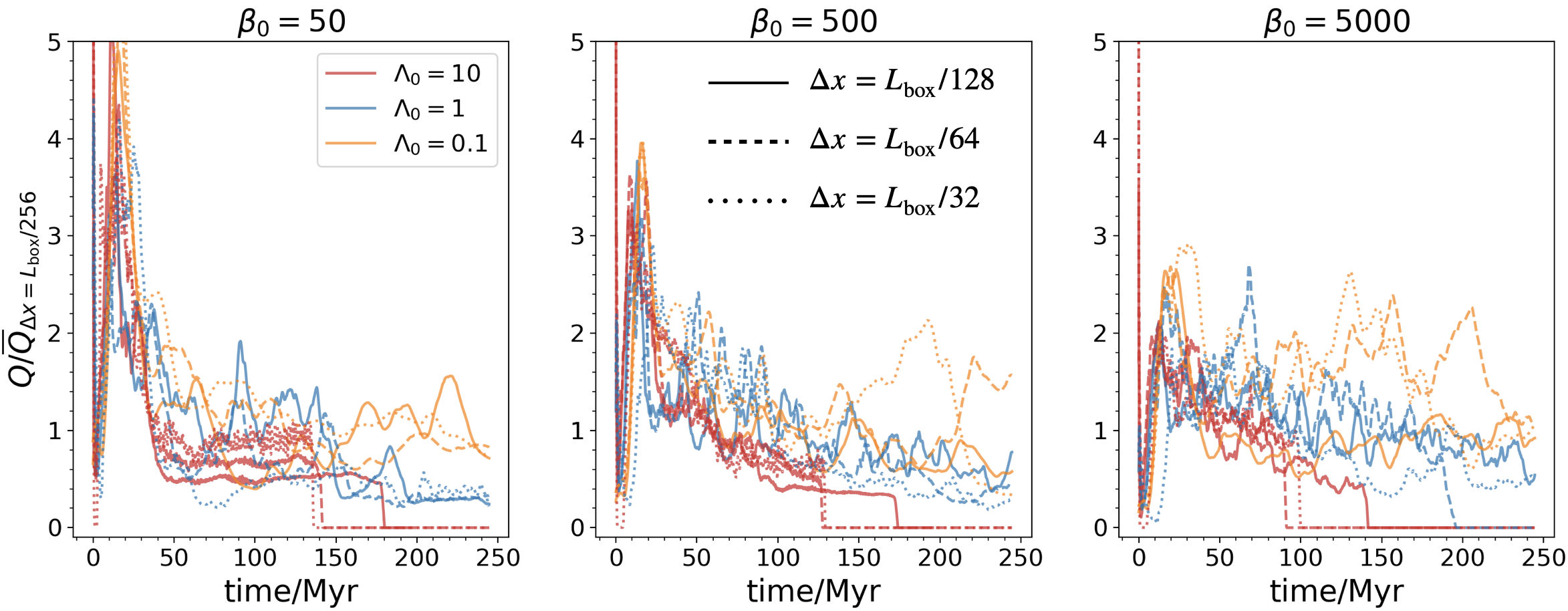}
\caption{\label{Q_append}Convergence test on the robustness of surface brightness $Q$. From left to right we show the time evolution of $Q$ in simulations with $\beta_0=50,\ 500\ {\rm and}\ 5000$. Different line styles represent different resolutions. We normalize the values of $Q$ by the corresponding average in our highest resolution runs $(\Delta x=L_{\rm box}/256)$.}

\end{center}
\end{figure*}

Since TMLs are intrinsically small scale structure hard to capture in large scale simulations, we assess whether the energy exchange across TMLs is well reflected there, by running simulations with resolutions 2 times $(\Delta x=L_{\rm box}/64)$ and 4 times worse $(\Delta x=L_{\rm box}/32)$ than our fiducial runs. The results are shown in Figure \ref{Q_append}. When cooling is weak, there can be inconsistency in certain period, but in term of average, the values of $Q$ are largely converged considering the level of fluctuations. Note that in a large scale simulation, such as the cloud-crushing problem, a typical resolution is $\Delta x=r_{\rm cloud}/64$, therefore our results inline with the view that, at late time, the local energy transfer across magnetized TMLs can be largely reproduced in a large scale simulation. However we stress that this conclusion must be further examined in the situation where magnetization is stronger and field geometry is more realistic.


\bsp	
\label{lastpage}
\end{document}